\documentclass[11pt]{scrartcl}

\usepackage{float}
\floatstyle{plaintop}
\restylefloat{table}
\usepackage[tableposition=top]{caption}

\usepackage[english]{babel}
\usepackage[utf8]{inputenc}
\usepackage[T1]{fontenc}
\usepackage[pdftex]{graphicx}
\usepackage{mathtools}
\usepackage{bbm}                
\usepackage{color}
\usepackage{txfonts}            
\usepackage{wasysym}            
\usepackage{xspace}
\usepackage{rotating}

\usepackage{lineno}

\newcommand{\GEVcc}{\mbox{$\mathrm{GeV}/{\textit{c}^2}$}}
\newcommand{\GEVc}{\mbox{$\mathrm{GeV}/{\textit{c}}$}}

\newcommand{\MEVcc}{\mbox{$\mathrm{MeV}/{\textit{c}^2}$}}
\newcommand{\MEVc}{\mbox{$\mathrm{MeV}/\textit{c}$}}
\newcommand{\MEV}{\mbox{$\mathrm{MeV}$}}
\newcommand{\KET}{\mbox{$K^\pm \to \pi^0 e^\pm \nu_e$}}
\newcommand{\KMT}{\mbox{$K^\pm \to \pi^0 \mu^\pm \nu_{\mu}$}}

\newcommand{\El}{E^\ast_l}

\newcommand{\Ep}{E^\ast_\pi}

\newcommand{\Elreco}{E^{\ast reco}_l}

\newcommand{\Epreco}{E^{\ast reco}_\pi}

\newcommand{\Epmax}{E^{\ast,\text{max}}_{\pi}}
\newcommand{\En}{E^\ast_\nu}
\newcommand{\lmp}{\lambda'_+}
\newcommand{\lmmp}{\lambda''_+}
\newcommand{\lmz}{\lambda_0}
\newcommand{\plnu}{\ensuremath{p_{\nu,\parallel}}\xspace}

\newcommand{\vus}{\ensuremath{|V_\text{us}|}\xspace}
\graphicspath{{./figures/}}

\begin{document}

\centerline{\LARGE EUROPEAN ORGANIZATION FOR NUCLEAR RESEARCH}
\vspace{15mm}
{\flushright{
CERN-EP-2018-231 \\
August 27, 2018 \\
 }}
\vspace{15mm}

\begin{center}
  \textsf{\textbf{\LARGE  Measurement of the form factors of \\*[4mm]
       charged kaon semileptonic decays }}  \\*[10mm]      
  \textsf{\textbf{\Large The NA48/2 Collaboration}}
\end{center}

\abstract{ A measurement of the  form factors of charged kaon semileptonic  decays  is presented,
based on   \mbox{$4.4 \times 10^6$}  $\KET$ ($K^\pm_{e3}$) and
\mbox{$2.3 \times 10^6$} $\KMT$ ($K^\pm_{\mu3}$) decays 
collected in 2004 by the NA48/2 experiment. The results  are obtained with improved precision 
as compared to earlier measurements. The combination of measurements in the 
$K^\pm_{e3}$ and $K^\pm_{\mu3}$ modes  is also presented.
}
\vspace{25mm}

\begin{center}
 {\em To be submitted for publication to JHEP}
\end{center}

\clearpage
\begin{center}
{\Large The NA48/2 Collaboration}\\
\vspace{2mm}
 J.R.~Batley,
 G.~Kalmus,
 C.~Lazzeroni$\,$\footnotemark[1]$^,$\footnotemark[2],
 D.J.~Munday$\,$\footnotemark[1],
 M.W.~Slater$\,$\footnotemark[1],
 S.A.~Wotton \\
{\em \small Cavendish Laboratory, University of Cambridge,
Cambridge, CB3 0HE, UK$\,$\footnotemark[3]} \\[0.2cm]
 R.~Arcidiacono$\,$\footnotemark[4],
 G.~Bocquet,
 N.~Cabibbo$\,$\renewcommand{\thefootnote}{\fnsymbol{footnote}}%
\footnotemark[2]\renewcommand{\thefootnote}{\arabic{footnote}},
 A.~Ceccucci,
 D.~Cundy$\,$\footnotemark[5], \\
 V.~Falaleev$\,$\footnotemark[6], 
 M.~Fidecaro, 
 L.~Gatignon,
 A.~Gonidec,
 W.~Kubischta, \\
 A.~Maier, 
 A.~Norton$\,$\footnotemark[7],
  M.~Patel$\,$\footnotemark[8],
 A.~Peters\\
{\em \small CERN, CH-1211 Gen\`eve 23, Switzerland} \\[0.2cm]
 S.~Balev$\,$\renewcommand{\thefootnote}{\fnsymbol{footnote}}
\footnotemark[2]\renewcommand{\thefootnote}{\arabic{footnote}},
 P.L.~Frabetti,
 E.~Gersabeck$\,$\footnotemark[9],
 E.~Goudzovski$\,$\footnotemark[1]$^,$\footnotemark[2]$^,$\footnotemark[10],
 P.~Hristov$\,$\footnotemark[11], \\
 V.~Kekelidze,
 V.~Kozhuharov$\,$\footnotemark[12]$^,$\footnotemark[13],
 L.~Litov$\,$\footnotemark[12],
 D.~Madigozhin\renewcommand{\thefootnote}{\fnsymbol{footnote}}
\footnotemark[1]\renewcommand{\thefootnote}{\arabic{footnote}},
  N.~Molokanova, \\
 I.~Polenkevich,
 Yu.~Potrebenikov,
 S.~Shkarovskiy\renewcommand{\thefootnote}{\fnsymbol{footnote}}
\footnotemark[1]\renewcommand{\thefootnote}{\arabic{footnote}},
 S.~Stoynev$\,$\footnotemark[14],
 A.~Zinchenko$\,$\renewcommand{\thefootnote}{\fnsymbol{footnote}}
\footnotemark[2]\renewcommand{\thefootnote}{\arabic{footnote}} \\
{\em \small Joint Institute for Nuclear Research, 141980 Dubna (MO), Russia} \\[0.2cm]
 E.~Monnier$\,$\footnotemark[15],
 E.~Swallow$\,$\renewcommand{\thefootnote}{\fnsymbol{footnote}}%
\footnotemark[2]\renewcommand{\thefootnote}{\arabic{footnote}},
 R.~Winston$\,$\footnotemark[16]\\
{\em \small The Enrico Fermi Institute, The University of Chicago,
Chicago, IL 60126, USA}\\[0.2cm]
 P.~Rubin$\,$\footnotemark[17],
 A.~Walker \\
{\em \small Department of Physics and Astronomy, University of Edinburgh, 
Edinburgh, EH9 3JZ, UK} \\[0.2cm]
 P.~Dalpiaz,
 C.~Damiani,
 M.~Fiorini,
 M.~Martini,
 F.~Petrucci,
 M.~Savri\'e,
 M.~Scarpa,
 H.~Wahl \\
{\em \small Dipartimento di Fisica e Scienze della Terra dell'Universit\`a e 
Sezione dell'INFN di Ferrara, \\
I-44122 Ferrara, Italy} \\[0.2cm]
 W.~Baldini,
 A.~Cotta Ramusino,
 A.~Gianoli\\
{\em \small Sezione dell'INFN di Ferrara,
I-44122 Ferrara, Italy} \\[0.2cm]
 M.~Calvetti,
 E.~Celeghini,
 E.~Iacopini,
 M.~Lenti,
 G.~Ruggiero$\,$\footnotemark[18] \\
{\em \small Dipartimento di Fisica dell'Universit\`a e Sezione
dell'INFN di Firenze, \\
I-50125 Sesto Fiorentino, Italy} \\[0.2cm]
 A.~Bizzeti$\,$\footnotemark[19],
 M.~Veltri$\,$\footnotemark[20] \\
{\em \small Sezione dell'INFN di Firenze, I-50019 Sesto Fiorentino, Italy} \\[0.2cm]
 M.~Behler,
 K.~Eppard,
 M.~Hita-Hochgesand,
 K.~Kleinknecht,
 P.~Marouelli,
 L.~Masetti, \\
 U.~Moosbrugger,
 C.~Morales Morales, 
 B.~Renk,
 M.~Wache,
 R.~Wanke,
 A.~Winhart$\,$\footnotemark[1] \\
{\em \small Institut f\"ur Physik, Universit\"at Mainz, D-55099
 Mainz, Germany$\,$\footnotemark[21]} \\[0.2cm]
 D.~Coward$\,$\footnotemark[22],
 A.~Dabrowski$\,$\footnotemark[11],
 T.~Fonseca Martin,
 M.~Shieh,
 M.~Szleper$\,$\footnotemark[23],\\
 M.~Velasco,
 M.D.~Wood$\,$\footnotemark[22] \\
{\em \small Department of Physics and Astronomy, Northwestern
University, Evanston, IL 60208, USA}\\[0.2cm]
 G.~Anzivino,
 E.~Imbergamo,
 A.~Nappi$\,$\renewcommand{\thefootnote}{\fnsymbol{footnote}}%
\footnotemark[2]\renewcommand{\thefootnote}{\arabic{footnote}},
 M.~Piccini,
 M.~Raggi$\,$\footnotemark[24],
 M.~Valdata-Nappi \\
{\em \small Dipartimento di Fisica dell'Universit\`a e
Sezione dell'INFN di Perugia, I-06100 Perugia, Italy} \\[0.2cm]
 P.~Cenci,
 M.~Pepe,
 M.C.~Petrucci \\
{\em \small Sezione dell'INFN di Perugia, I-06100 Perugia, Italy} \\[0.2cm]
 F.~Costantini,
 N.~Doble,
 L.~Fiorini$\,$\footnotemark[25],
 S.~Giudici,
 G.~Pierazzini$\,$\renewcommand{\thefootnote}{\fnsymbol{footnote}}%
\footnotemark[2]\renewcommand{\thefootnote}{\arabic{footnote}},
 M.~Sozzi,
 S.~Venditti  \\
{\em Dipartimento di Fisica dell'Universit\`a e Sezione dell'INFN di
Pisa, I-56100 Pisa, Italy} \\[0.2cm]
 G.~Collazuol$\,$\footnotemark[26],
 L.~DiLella$\,$\footnotemark[27],
 G.~Lamanna$\,$\footnotemark[27],
 I.~Mannelli,
 A.~Michetti \\
{\em Scuola Normale Superiore e Sezione dell'INFN di Pisa, I-56100
Pisa, Italy} \\[0.2cm]
 C.~Cerri,
 R.~Fantechi \\
{\em Sezione dell'INFN di Pisa, I-56100 Pisa, Italy} \\[0.2cm]
\clearpage
 B.~Bloch-Devaux$\,$\footnotemark[28],
 C.~Cheshkov$\,$\footnotemark[29],
 J.B.~Ch\`eze,
 M.~De Beer,
 J.~Derr\'e, \\
 G.~Marel, 
 E.~Mazzucato,
 B.~Peyaud,
 B.~Vallage \\
{\em \small DSM/IRFU -- CEA Saclay, F-91191 Gif-sur-Yvette, France} \\[0.2cm]
%
 M.~Holder,
 M.~Ziolkowski \\
{\em \small Fachbereich Physik, Universit\"at Siegen, D-57068
 Siegen, Germany$\,$\footnotemark[30]} \\[0.2cm]
 S.~Bifani$\,$\footnotemark[1],
 M.~Clemencic$\,$\footnotemark[11],
 S.~Goy Lopez$\,$\footnotemark[31] \\
{\em \small Dipartimento di Fisica dell'Universit\`a e
Sezione dell'INFN di Torino,\\ I-10125 Torino, Italy} \\[0.2cm]
 C.~Biino,
 N.~Cartiglia,
 F.~Marchetto \\
{\em \small Sezione dell'INFN di Torino, I-10125 Torino, Italy} \\[0.2cm]
 H.~Dibon,
 M.~Jeitler,
 M.~Markytan,
 I.~Mikulec,
 G.~Neuhofer,
 L.~Widhalm$\,$\renewcommand{\thefootnote}{\fnsymbol{footnote}}%
\footnotemark[2]\renewcommand{\thefootnote}{\arabic{footnote}} \\
{\em \small \"Osterreichische Akademie der Wissenschaften, Institut
f\"ur Hochenergiephysik,\\ A-10560 Wien, Austria$\,$\footnotemark[32]} \\[0.5cm]
\end{center}

\setcounter{footnote}{0}
\renewcommand{\thefootnote}{\fnsymbol{footnote}}
\footnotetext[1]{Corresponding authors, email:  Dmitri.Madigojine@cern.ch, Sergey.Shkarovskiy@cern.ch}
\footnotetext[2]{Deceased}
\renewcommand{\thefootnote}{\arabic{footnote}}
\footnotetext[1]{Now at: School of Physics and Astronomy, University of Birmingham,  Birmingham, B15 2TT, UK}
\footnotetext[2]{
Supported by a Royal Society University Research Fellowship
(UF100308, UF0758946)}
\footnotetext[3]{Funded by the UK Particle Physics and Astronomy Research Council, grant PPA/G/O/1999/00559}
\footnotetext[4]{Now at: Universit\`a degli Studi del Piemonte Orientale e Sezione 
dell'INFN di Torino, I-10125 Torino, Italy}
\footnotetext[5]{Now at: Istituto di Cosmogeofisica del CNR di Torino,
I-10133 Torino, Italy}
\footnotetext[6]{Now at: Joint Institute for Nuclear Research, 141980 Dubna (MO), Russia}
\footnotetext[7]{Now at: Dipartimento di Fisica e Scienze della Terra dell'Universit\`a e Sezione
dell'INFN di Ferrara, I-44122 Ferrara, Italy}
\footnotetext[8]{Now at: Department of Physics, Imperial College, London,
SW7 2BW, UK}
\footnotetext[9]{Now at: School of Physics and Astronomy, The University of Manchester, Manchester, M13 9PL, UK}
\footnotetext[10]{Supported by ERC Starting Grant 336581}
\footnotetext[11]{Now at: CERN, CH-1211 Gen\`eve 23, Switzerland}
\footnotetext[12]{Now at: Faculty of Physics, University of Sofia ``St. Kl.
Ohridski'', BG-1164 Sofia, Bulgaria, 
funded by the Bulgarian National Science Fund under contract DID02-22}
\footnotetext[13]{Also at: Laboratori Nazionali di Frascati, I-00044 Frascati, Italy}
\footnotetext[14]{Now at: Fermi National Accelerator Laboratory, Batavia, IL 60510, USA}
\footnotetext[15]{Now at: Centre de Physique des Particules de Marseille,
IN2P3-CNRS, Universit\'e de la M\'editerran\'ee, F-13288 Marseille,
France}
\footnotetext[16]{Now at: School of Natural Sciences, University of California, Merced, CA 95343, USA}
\footnotetext[17]{Now at: School of Physics, Astronomy and Computational Sciences, George Mason
University, Fairfax, VA 22030, USA}
\footnotetext[18]{Now at: Physics Department, University of Lancaster, Lancaster, LA1 4YW, UK} 
\footnotetext[19]{Also at Dipartimento di Scienze Fisiche, Informatiche e Matematiche, Universit\`a di Modena e Reggio Emilia, I-41125 Modena, Italy}
\footnotetext[20]{Also at Istituto di Fisica, Universit\`a di Urbino,
I-61029 Urbino, Italy}
\footnotetext[21]{Funded by the German Federal Minister for
Education and research under contract 05HK1UM1/1}
\footnotetext[22]{Now at: SLAC, Stanford University, Menlo Park, CA 94025, USA}
\footnotetext[23]{Now at: National Center for Nuclear Research, P-05-400 \'Swierk, Poland}
\footnotetext[24]{Now at: Universit\`a di Roma ``La Sapienza'', I-00185 Roma, Italy}
\footnotetext[25]{Now at: Instituto de F\'{\i}sica Corpuscular IFIC,
Universitat de Val\`{e}ncia, E-46071 Val\`{e}ncia, Spain}
\footnotetext[26]{Now at: Dipartimento di Fisica dell'Universit\`a e Sezione dell'INFN di Padova, I-35131 Padova, Italy}
\footnotetext[27]{Now at: Dipartimento di Fisica dell'Universit\`a e Sezione dell'INFN di Pisa, I-56100 Pisa, Italy}
\footnotetext[28]{Now at: Dipartimento di Fisica dell'Universit\`a di Torino, 
I-10125 Torino, Italy}
\footnotetext[29]{Now at: Institut de Physique Nucl\'eaire de Lyon,
IN2P3-CNRS, Universit\'e Lyon I, F-69622 Villeurbanne, France}
\footnotetext[30]{Funded by the German Federal Minister for Research
and Technology (BMBF) under contract 056SI74}
\footnotetext[31]{Now at: Centro de Investigaciones Energeticas
Medioambientales y Tecnologicas, E-28040 Madrid, Spain}
\footnotetext[32]{Funded by the Austrian Ministry for Traffic and
Research under the contract GZ 616.360/2-IV GZ 616.363/2-VIII, and
by the Fonds f\"ur Wissenschaft und Forschung FWF Nr.~P08929-PHY}


\clearpage

\section{Introduction}
\label{intro}

The NA48/2 experiment at the CERN SPS was designed primarily to search
for direct CP violation in $K^\pm$ decays to three
pions~\cite{Batley:2007aa}. It used simultaneous $K^+$ and
$K^-$ beams with momenta of $60~\GEVc$. 
Data were collected in  2003--2004, providing   $2\times 10^9$ 
reconstructed $K^\pm \to 3\pi$ decays.  Additionally, a  data set  was recorded at reduced beam intensity  using a minimum bias trigger during a 52-hour long data-taking period in 2004.

The $K^\pm\to\pi^0 l^\pm\nu$ ($K^{\pm}_{l3}$, with $l=e,\mu$)  decays  contribute to the 
precise determination of the CKM matrix element \vus \cite{Antonelli:2010yf},
 which requires the knowledge of both branching ratios and form factors (FFs). 
 Measurements of  the $K^{\pm}_{l3}$ vector $f_+$ and scalar  $f_0$  FFs 
 based on the above minimum bias data set are presented  here.

In absence of electromagnetic effects, the
 differential $K^\pm _{l3}$ decay rate  is described in the ($\El,\Ep$) Dalitz plot as~\cite{Chounet:1971yy}:
\begin{equation}
  \frac{d^2 \, \Gamma(K^\pm _{l3})}{d \El \, d \Ep} \: \:
    = \: \rho(\El,\Ep) \:
         = \: N \left( \, A_1 \, |f_+ (t) |^2 + A_2 \, f_+(t)f_- (t) + A_3 \, |f_- (t) |^2 \, \right),
  \label{dalitz0}
\end{equation}
where $\El$ and $\Ep$ are the lepton and pion energies in the kaon rest frame;
$t$ is the 4-momentum transfer to the leptonic system;
$N$ is a  numerical factor;  $f_-(t) = (f_0(t) - f_+(t))(m_K^2 -m_{\pi^0}^2)/t$;
$m_K$  and $m_{\pi^0}$ are the
charged kaon and neutral pion masses~\cite{Olive:2016xmw}. The kinematic factors are 
\begin{align}
  A_1 &\: = \:  m_K \, \left(2\,\El\,\En - m_K(\Epmax - \Ep)\right) + m_l^2 \, \left((\Epmax - \Ep)/4 - \En\right), \\ 
  A_2 &\: = \:  m_l^2 \, \left(\En - (\Epmax - \Ep)/2\right) \nonumber, \\ 
  A_3 &\: = \:  m_l^2 \, (\Epmax - \Ep)/4.  \nonumber
\end{align}
Here $\Epmax = (m_K^2 + m_{\pi^0}^2 - m_l^2)/2 \, m_K$,  $m_l$ is the charged lepton mass, 
and $\En = m_K - \El - \Ep$ is the neutrino energy in the kaon rest frame. 
For $K^\pm _{e3}$ decays, the factors $A_2$ and $A_3$, which are proportional to $m_l^2 $, become negligible and only the vector FF  contributes within the experimental precision.

The   FF parameterizations  considered  are described in Table~\ref{parameterizations}.
They include a Taylor expansion in the variable  $t/m_{\pi^+}^2 $ \cite{Olive:2016xmw}, where $m_{\pi^+}$ is the charged pion mass,
a parameterization assuming vector and scalar pole masses $M_V$ and $M_S$~\cite{Dennery:1963, Lichard:1997ya} and 
a more physical dispersive parameterization~\cite{Bernard:2009zm}. 
The Taylor expansion is affected by large  correlations between the measured parameters.  
The pole parameterization  has a physical interpretation for $f_+ (t) $  
related to the $K^\ast(892)$ scattering pole, but not  for  $f_0 (t) $  with no corresponding pole.
The dispersive parameterization makes use of general  chiral symmetry and analyticity constraints, and  external inputs from $K$--$\pi$ scattering data, via 
the functions $H(t)$ and $G(t)$: 
\begin{equation}\label{eq:gh}
  \begin{aligned}
    G(t) &\: = \: x \cdot G_{p1} + (1-x) \cdot G_{p2} + x \cdot (1-x) \cdot G_{p3}, \\ 
    H(t) &\: = \: x \cdot H_{p1} + x^2 \cdot H_{p2}, 
  \end{aligned}
\end{equation}
 with $x=t/(m_K-m_{\pi^0})^2$, and  the numerical values of the parameters~\cite{Bernard:2009zm}:
\begin{equation}
\begin{aligned}
  G_{p1} &=  0.0209 \pm 0.0021,  ~~G_{p2} = 0.0398 \pm 0.0044,  ~~G_{p3} =  0.0045 \pm 0.0004,   \\
   H_{p1} &= (1.92^{+0.63}_{-0.32})\cdot 10^{-3},  
   ~~H_{p2} = (2.63^{+0.28}_{-0.15})\cdot 10^{-4}.
 \end{aligned}
\end{equation}

\begin{table}[hbt]
  \renewcommand{\arraystretch}{1.2}
  \centering
  \begin{tabular}{|c|c|c|} \hline
                     & $f_+(t)$ & $f_0(t)$ \\*[0.2em]
    \hline  & & \\*[-1.2em]
    Taylor expansion & $\displaystyle 1 + \lmp \, \frac{t}{m_{\pi^+}^2} + \tfrac{1}{2} \, \lmmp \, \left(\frac{t}{m_{\pi^+}^2}\right)^2$
                     & $\displaystyle 1 + \lmz \, \frac{t}{m_{\pi^+}^2}$ \\*[1em]
    \hline  & & \\*[-1.2em]
    Pole             & $\displaystyle \frac{M_V^2}{M_V^2-t}$ 
                     & $\displaystyle \frac{M_S^2}{M_S^2-t}$ \\*[1em] 
    \hline & & \\*[-1.2em]
    Dispersive       & $\displaystyle \exp{\left(\frac{\Lambda_+ + H(t)}{m_{\pi^+}^2} \, t \right)}$
                     & $\displaystyle \exp{\left(\frac{\ln{C}-G(t)}{m_K^2 -m_{\pi^0}^2}  \, t\right)}$ \\*[1em] \hline
  \end{tabular}
  \caption{Form factor parameterizations used in this analysis. 
  The free parameters to be measured are the $\lmp$, $\lmmp$, $\lmz$ coefficients (slopes) for the Taylor expansion, 
       the scalar $M_S$ and vector $M_V$ mass values for the pole model, and the $\Lambda_+$ and $\ln{C}$ parameters for the dispersive model.}
  \label{parameterizations}
  \renewcommand{\arraystretch}{1.0}
\end{table}

\section{Beams and detectors}
\label{sec:beamdet}
Detailed descriptions of the NA48/2 beam line and detectors  are
available in Refs.~\cite{Batley:2007aa, Fanti:2007vi}.
 Two simultaneous charged hadron beams 
produced by $400~\GEVc$ protons impinging on a beryllium target were used.  Kaons represented  6\%  of the total beam flux and the  $K^+ /K^- $ flux ratio was 1.79.
Particles of opposite charge with a central momentum of $60~\GEVc$ and
a momentum band of $\pm 3.8\%$ (RMS) 
were selected by a system of dipole magnets, focusing quadrupoles,
muon sweepers and collimators. 
The decay volume was contained in a 114~m long vacuum tank with a diameter of
1.92~m for the first 66~m, and 2.40~m  downstream.
The two beams were superimposed in the decay volume along a common axis  which defined 
the Z axis  of the coordinate system. The Y axis 
pointed vertically up, and the X axis was directed horizontally to form a right-handed system.

Charged particles from $K^\pm$ decays were measured by a magnetic
spectrometer  consisting of four drift chambers
(DCH1--DCH4) and a dipole magnet between DCH2 and DCH3. Each chamber
 consisted of four staggered double planes of sense wires 
measuring the  coordinates transverse to the beam axis along the 0$^\circ$, 90$^\circ$ and $\pm\,$45$^\circ$ directions.
The spectrometer was located in a tank filled with helium at nearly
atmospheric pressure  and separated from the vacuum tank
by a 0.3\% $X_0$ thick \textit{Kevlar}\textsuperscript{\textregistered} 
window.  
A  15.8~cm diameter evacuated aluminium tube traversing the centre of the main  detectors 
allowed the undecayed beam particles and the muon halo from beam pion decays to continue their path in vacuum.
The spectrometer momentum resolution was
$\sigma_p /p=1.02\% \oplus 0.044\% \cdot p$, with the momentum $p$ expressed in $\GEVc$.
The spectrometer was followed by a scintillator hodoscope
(HOD) consisting of two planes segmented into horizontal and vertical
strips and arranged in four quadrants.

A liquid krypton calorimeter (LKr) was used to
reconstruct $\pi^0 \to \gamma \gamma$ decays and for charged particle identification. It  is a
 27~$X_0$ thick 
 quasi-homogeneous ionization chamber with an active volume of  
7 m$^3$ of liquid krypton,  segmented transversally  into 13248  $2 \times 2$~cm$^2$  projective cells.
 It provided an energy resolution $\sigma_E /E = 0.032/\sqrt{E} \oplus 0.09/E \oplus 0.0042$, 
a   resolution on the transverse coordinates of an isolated electromagnetic shower 
$\sigma_x = \sigma_y = (0.42/\sqrt{E} \oplus 0.06)$~cm,  
and  a time resolution   $\sigma_t = (2.5/\sqrt{E})$~ns, with $E$ expressed in GeV.
A hodoscope (NHOD) consisting of a plane of scintillating
fibers,  located inside the LKr calorimeter,  was used for triggering purposes.  

The LKr was followed by a hadronic calorimeter with a total iron
thickness of 1.2~m. A muon detector (MUV), located further
downstream, consisted of three planes of 2.7~m long and 2~cm thick scintillator strips (28~strips in total)
read out by photomultipliers at both ends. Each plane was
preceded by a 80~cm thick iron wall. The strips were aligned
horizontally in the first and the last planes, and vertically in the second plane. 

During the considered data-taking period,  $4.8 \times 10^8$  events were recorded using a
minimum bias trigger  condition requiring  a coincidence of signals in the two HOD planes in the same quadrant and  an energy deposit above 10~GeV in the LKr. 
The data set  is divided into twelve sub-samples according to the polarities of the beam 
line and spectrometer magnets that interchanged the paths of the positive and negative  beams.

\section{Monte Carlo simulation}
\label{MCsim}

A GEANT3-based~\cite{Brun:1978fy} Monte Carlo (MC) simulation including beam line, detector geometry and material description is used to evaluate the detector response. The beam simulation is tuned using the kaon momentum and direction distributions as measured from reconstructed $K^\pm\to\pi^\pm\pi^+\pi^-$ decays. MC samples of $K^\pm _{e3}$ ($K^\pm _{\mu3}$) decays corresponding to 3 (5) times the data samples have been produced.

The $K^\pm _{l3}$ decays are modelled according to~\cite{Gatti:2005kw} including both the Dalitz plot density of  Eq.~(\ref{dalitz0}) and radiative corrections, with  exactly one   photon emitted in each decay, and tracked through the detector if its energy in the laboratory frame is above 1~MeV. 
 This approach  takes into account the infrared divergence of photon radiation by extending the soft-photon approximation~\cite{Weinberg:1965nx} to the whole energy range. 
The implementation has been validated  in~\cite{Gatti:2005kw} using the experimental data available  at the time \cite{KTEV,NA48}: photon energy and photon-lepton angle distributions have been found to agree with the data within 1--5\% systematic uncertainty. However this uncertainty includes the effect of a 100\% variation of the vector FF slope. Therefore the distributions considered are not sensitive to the FF description at the level of precision required for the present study.  

On the other hand, model-independent (universal) radiative corrections have been proposed  in~\cite{Cirigliano:2001mk}. 
Using these corrections, the effects of model- and approximation-dependent interplay between QED and QCD are  absorbed in the measured effective FFs.
 These FFs are free from uncertainties due to radiative corrections  by construction,  and
 their deviation  from FFs defined in absence of electromagnetic interaction can be estimated within the formalism used by~\cite{Cirigliano:2001mk}. However this approach does not include real photon emission.

In this analysis, the approach of~\cite{Gatti:2005kw} is used, and  the Dalitz plot density is corrected by event-by-event weights $w_r(\El,\Ep)$ equal to the ratio of densities obtained within the formulations of \cite{Cirigliano:2001mk} and \cite{Gatti:2005kw}.
In the $K^\pm _{e3}$ case, the weighting leads to $d\Gamma / dE^*_e$ variations 
as large as 2\%.
In the $K^\pm _{\mu3}$ case, the weights  have been found to be $w_r(E^* _\mu, \Ep) = 1$ within the required precision. 
A linear approximation for the vector and scalar FFs $f_+ (t) = f_0 (t) = 1 + ~0.0296 \cdot t / m_{\pi ^+} ^2$  is used to generate the simulated samples.

\section{Event  selection and reconstruction}
\label{sec:selection}
Charged particles (trajectories and momenta) and LKr energy deposition clusters (energies and positions) are reconstructed  as described in~\cite{Batley:2007aa}.  The energy scale correction applied to LKr clusters is  established from a  study of the energy-to-momentum ratio of reconstructed electrons. 
\subsection{Neutral pion selection}\label{sec:pi0}
Photon candidates are defined as LKr clusters satisfying the following requirements:
energy above 3~GeV;
distances to impact points at the  LKr front plane of each in-time (within $\pm$10~ns) track
 larger than 15~cm;  distances to other in-time  (within $\pm$5~ns) clusters larger than 10~cm.
In addition,  photon candidates are required to be at least 8 cm away from the LKr edges and 2 cm away  from each of the 49 inactive cells  to reduce the effects of energy losses.

A pair of in-time (within $\pm$5~ns) photon candidates  is considered as a $\pi^0 \to \gamma\gamma$ decay candidate 
if there  are no additional photon candidates within $\pm$5~ns  of their average time,
the distance between them is  larger than  20~cm, and the sum of their energies is at least 15~GeV.  
The latter condition  ensures a high trigger efficiency. 

\begin{figure}[t]
\begin{center} 
\includegraphics[width=0.86\linewidth]{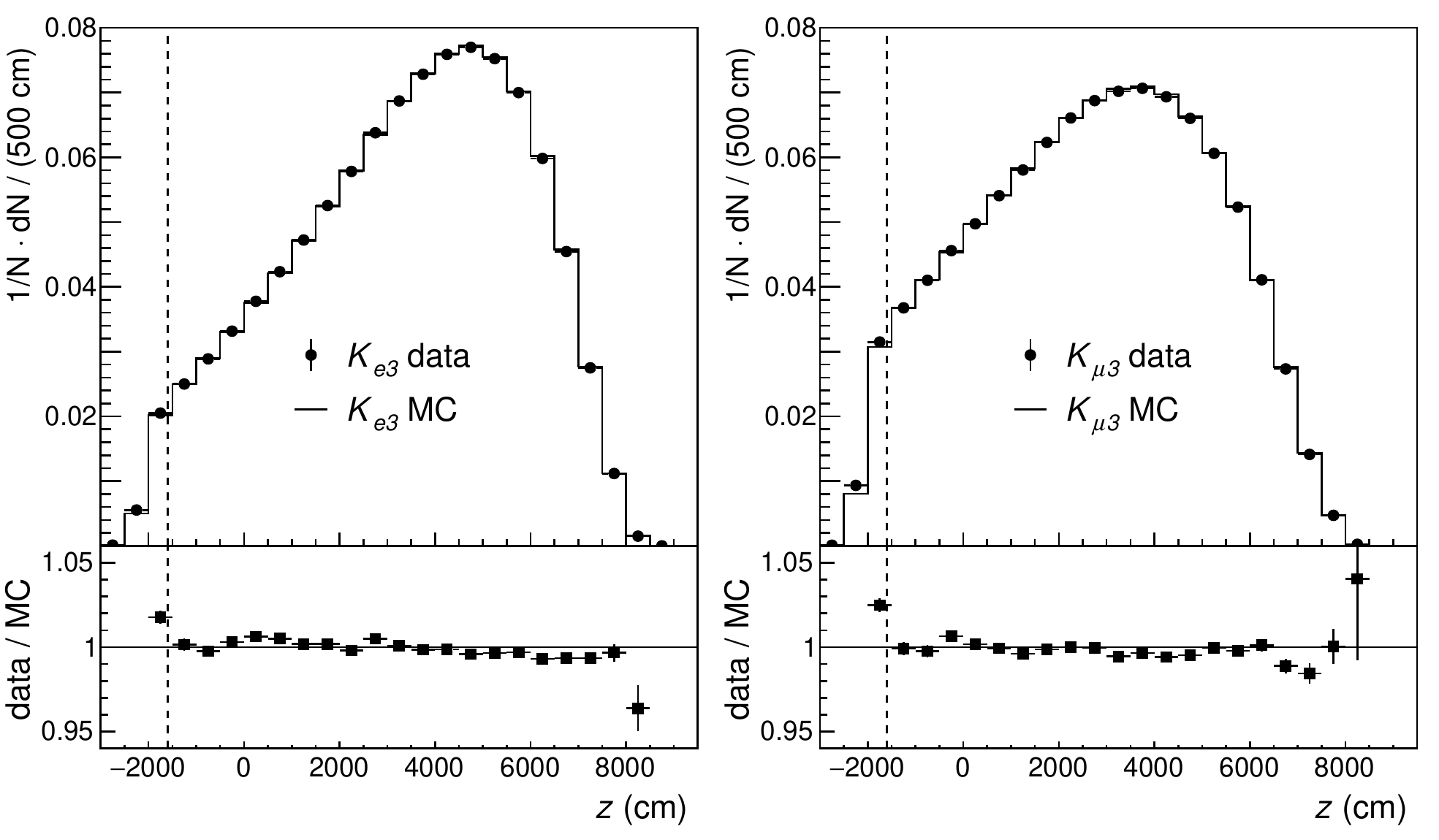}
\end{center}
\caption{Distributions of the decay vertex $z$ position for data  and MC  simulated samples for $K^\pm _{e3}$ (left) and   $K^\pm _{\mu3}$ (right) modes and corresponding  Data/MC ratios. The simulated samples include signal and backgrounds.  The vertical dashed lines indicate  the cut applied (the final collimator exit is located at $-1800$~cm).} 
 \label{fig:acceptance}
\end{figure}

The $z$ position of the $\pi^0 \to \gamma\gamma$ decay vertex is computed from photon candidate positions and energies assuming the nominal $\pi^0$ mass~\cite{Olive:2016xmw}.
It is required to be at least 2~m downstream of the final beam collimator to suppress 
$\pi^0$ production  in the material of the collimator (Fig.~\ref{fig:acceptance}). In addition, photons are 
required not to  intercept DCH beam pipe flanges~\cite{Batley:2000zz}.

\subsection{Charged lepton selection}

Lepton candidates are defined as reconstructed DCH tracks satisfying the following requirements.
Their momentum should be at least  5 (10)~$\GEVc$ for  $e^\pm$ ($\mu^\pm$) candidates,  the latter ensuring high  muon identification efficiency. 
The distance from the track impact point  at the LKr front  plane to the closest  inactive cell  should
exceed 2~cm, and the distance to the Z axis  in each DCH plane should be at least 15~cm. 
The track  should be in time (within $\pm$10~ns) with a $\pi^0$ candidate, and no additional tracks  are allowed within $\pm$8~ns of  the track. 

Tracks with  the ratio of LKr energy deposit $E$  to momentum $p$ in the range $0.9 < E/p < 2.0$  are identified as electrons ($e^\pm$).  Tracks with $E/p < 0.9$ and associated signals in the first two MUV planes are identified as muons.
Extrapolated muon track positions  at the first MUV plane
are required to be at least 30 (20)~cm away from the Z axis  (detector outer edges) 
 to reduce geometrical inefficiencies due to multiple scattering in the preceding material.

The $K^\pm _{l3}$ decay vertex is defined as follows: its $z$ coordinate is that of the $\pi^0$ decay (Section \ref{sec:pi0}), and its transverse $(x,y)$ coordinates are those of the  lepton track at this $z$ plane.
\subsection{Beam  profiles}
\label{sec:decayvertex}

The specific beam conditions of the data sample  triggered further studies of the transverse beam profiles  with  fully reconstructed 
$K^\pm \to \pi^\pm\pi^+\pi^-$ decays. These studies showed evidence for a diverging  beam component
surrounding the core  and  giving rise to kaon decay vertices a few centimetres off 
the Z axis. This component, which is likely to arise from  quasi-elastic kaon scattering  in 
 the beam line, is described using the following variable:
\begin{equation}
B = \sqrt{\,\left(\frac{x-x_0(z)}{\sigma_{x}(z)}\right)^2 
    + \left(\frac{y-y_0(z)}{\sigma_{y}(z)}\right)^2},
\end{equation}
where $x,y,z$  are the $K^\pm _{l3}$ decay vertex coordinates, 
 $x_0(z)$, $y_0(z)$ are the measured central positions of the beam profiles at the  vertex $z$ position, 
and $\sigma_{x}(z)$, $\sigma_{y}(z)$  are their Gaussian widths 
which decrease from 1 cm at the beginning to 0.6 cm at the end of the decay volume.   The beam profile  characteristics are obtained from  reconstructed $K^\pm \to \pi^\pm\pi^+\pi^-$ decays. 

\begin{figure}[t]
  \begin{center}
    \includegraphics[width=0.94\linewidth]{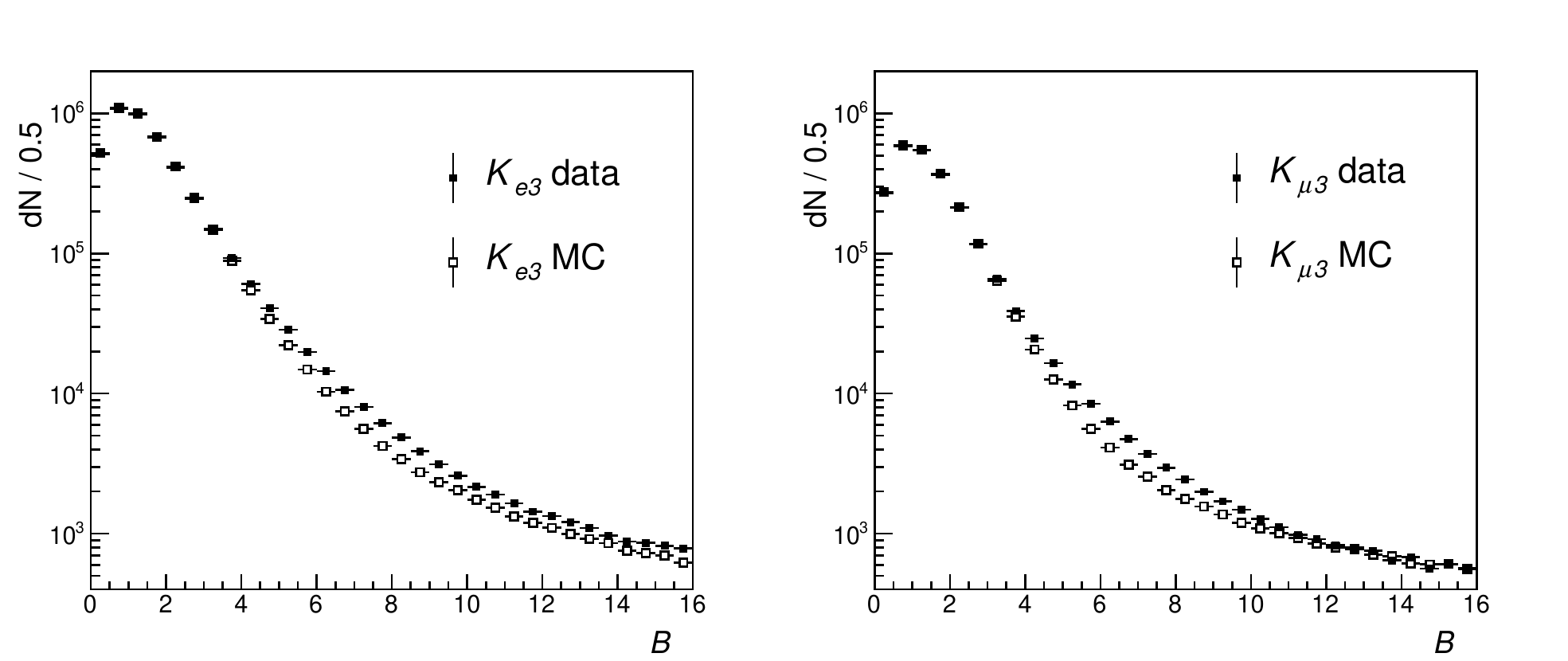}
  \end{center}
  \caption{Distributions of the beam  variable   $B$ for $K^\pm _{e3}$ (left) and $K^\pm _{\mu3}$ (right) 
      for data  and normalized MC samples. The simulated samples  include signal and backgrounds.     
  }
  \label{fig:Bell}
\end{figure}

The $B$ distributions of data and MC simulated events  are shown in Fig.~\ref{fig:Bell}. 
The data distributions are well described  by simulation in the core region ($B < 3$),  
while the diverging beam component in the data, which is not simulated, can be seen at larger $B$ values.
Quasi-elastic scattering affects marginally the kaon  momentum magnitude.  Scattered
beam kaons are conservatively considered in the analysis  by requiring $B < 11$, which minimizes the effect of correlations between  kaon  directions and momenta.  This condition also reduces  the background from $\pi^\pm$ decays in flight (Section~\ref{BGsection}).

\subsection{Kaon and neutrino momenta reconstruction}
\label{kaneumom}

\begin{figure}[p]
  \begin{center}
        \includegraphics[width=0.97\linewidth]{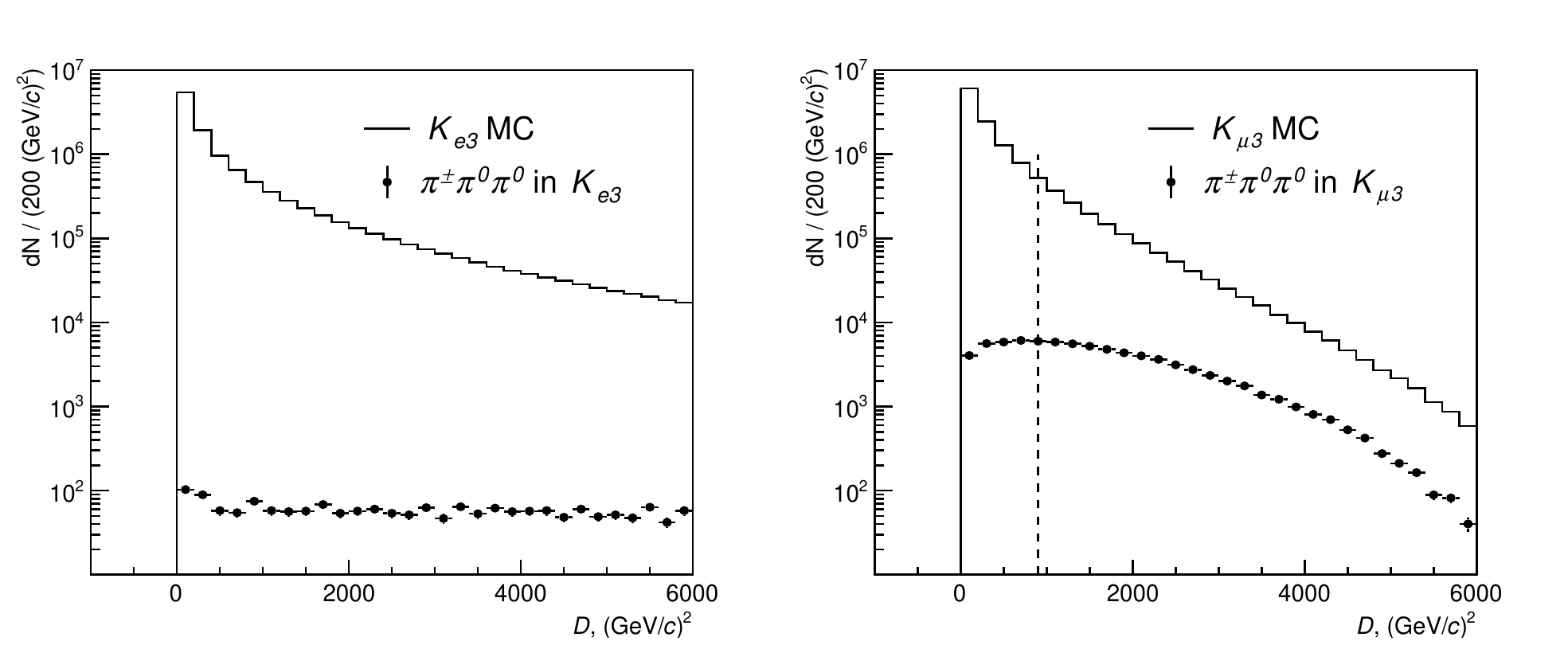}
  \end{center}
  \caption{ Distributions of the reconstructed $D$ variable  for MC simulated $K^\pm _{e3}$ (left) and 
  $K^\pm _{\mu3}$ (right) signal  and $K^\pm \to \pi^\pm\pi^0\pi^0$ background  samples. 
    The    selection condition
   $D <  900~(\GEVc)^2$, applied in the $K^\pm _{\mu3}$ case for background suppression, is indicated by the vertical dashed line.}
  \label{fig:dcut}
\end{figure}

A more precise estimate of the $K^\pm$ momentum magnitude ($p_K$) in the laboratory frame than the beam average value is obtained by imposing energy-momentum conservation in the kaon decay under the assumption of a missing neutrino, and fixing the kaon mass to its nominal value and the kaon direction to the measured beam axis direction. This leads to two solutions:
\begin{equation}
  p_K       =  \frac{\psi \, p_\parallel}{E^2 - p_\parallel^2} \pm \sqrt{D},
 \label{eq:pk}
\end{equation}
\begin{equation}
  \text{where} ~~~\psi    =   \tfrac{1}{2} \, ( m_K^2 + E^2 - p_\perp^2 - p_\parallel^2),  
    ~~~D =   \frac{\psi^2 \, p_\parallel^2}{(E^2 - p_\parallel^2)^2} - \frac{m_K^2 \, E^2 - \psi^2}{E^2 - p_\parallel^2}.
\end{equation}
 If $D$ is negative due to resolution effects, a value $D=0$ is used in the calculation.
Here  $E$,  $p_\parallel$ and $p_\perp$ are the energy, longitudinal and transverse momentum components  (with respect to the beam axis)  of the $\pi^0 l^\pm$ system in the laboratory frame.
The distributions of the $D$ variable for MC simulated events are shown in Fig.~\ref{fig:dcut}.  
 The solution that is closer to  the average beam momentum $p_B$ (measured from 
 $K^\pm \to \pi^\pm \pi^+ \pi^-$ decays) is chosen,  
and  required to satisfy  $|p_K - p_B| <  7.5~\GEVc$. 

\begin{figure}[p]
  \begin{center}
    \includegraphics[width=0.88\linewidth]{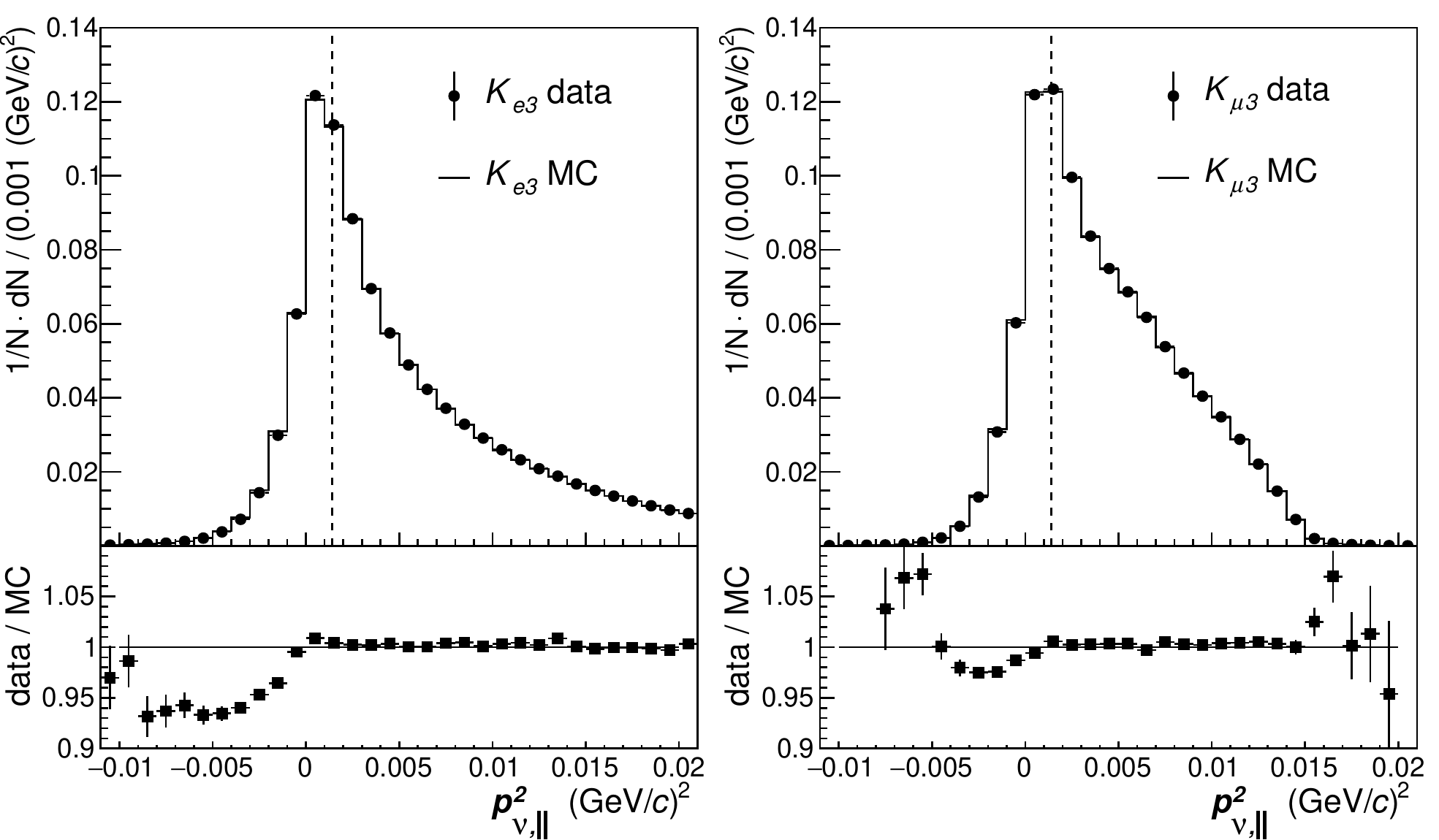}
  \end{center}
  \caption{Normalized $\plnu^2$ distributions of data and MC simulated   samples  for   $K^\pm _{e3}$ 
  (left) and $K^\pm _{\mu3}$ (right) modes and corresponding Data/MC ratios.  The simulated samples 
  include signal and backgrounds. The vertical dashed lines indicate the  $\plnu^2> 0.0014~(\GEVc)^2$ 
   cut applied.}   \label{fig:pnu2neum}
\end{figure}

Distributions of the squared neutrino longitudinal momentum in the kaon rest frame,
$\plnu^2 =  (m_K-E^\ast)^2-p_\perp^2$,  where $E^\ast$  is the $\pi^0 l^\pm$ system energy in the kaon 
rest frame, are shown in Fig.~\ref{fig:pnu2neum}.  The simulated spectra are  sensitive to details of the 
beam geometry description at small  $\plnu^2$  values, and negative values originate from resolution 
effects.  To ensure good agreement of data and simulation, it is required that
$\plnu^2> 0.0014~(\GEVc)^2$ (corresponding to  $\plnu > 37.4~\MEVc$)
which rejects 29\%  of the $K^\pm _{l3}$ events in  both decay modes. 
 
\subsection{Background suppression}
\label{BGsection}

 The $K^\pm \to \pi^\pm\pi^0\pi^0$ $(\pi^0 \to \gamma \gamma, ~\pi^0 \to \gamma \gamma)$ decays  contribute to the background if one of the $\pi^0$ mesons
is not detected, and  the $\pi^\pm$ either decays or is  misidentified.
This background affects mainly  the $K^\pm _{\mu3}$  sample, and  is  reduced by requiring  $D < 900~(\GEVc)^2$ in this case, as illustrated in Fig.~\ref{fig:dcut}.

The $K^\pm \to \pi^\pm \pi^0$ background in the $K^\pm _{e3}$  sample  arising from $\pi^\pm$ misidentification  is characterized by small total transverse momentum and is reduced  by requiring 
$p_{\nu,\perp} > 30~\MEVc$,  
taking into account resolution and beam  divergence effects.

\begin{figure}[tb]
  \begin{center}
    \includegraphics[width=0.90\linewidth]{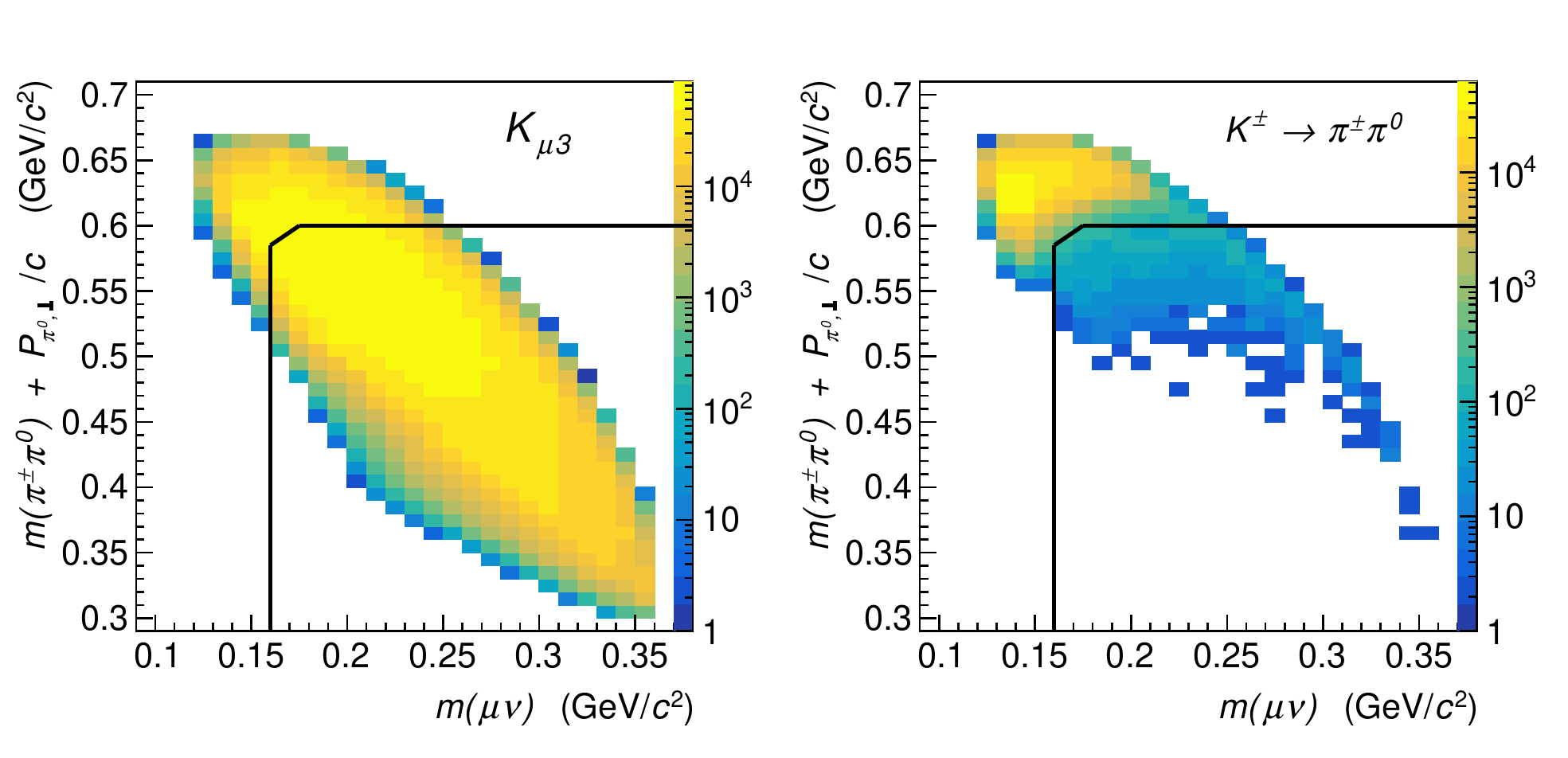}
  \end{center}
  \caption{Distributions of the kinematic variables used for $K^\pm \to \pi^\pm \pi^0$ background suppression for MC simulated  signal $K^\pm _{\mu3}$ (left) and background $K^\pm \to \pi^\pm\pi^0$ (right) samples. The selection criteria are indicated by solid lines.
  }
  \label{fig:km3cuts}
\end{figure}

The $K^\pm \to \pi^\pm \pi^0$ background to $K^\pm_{\mu 3}$  decays arises from  
$\pi^\pm$ misidentification and $\pi^\pm \to \mu^\pm \nu$ decay. 
 The former process  is suppressed by requiring the $\pi^0 l^\pm$ mass, reconstructed in the $\pi^+$ mass hypothesis for the lepton candidate, 
 to be $m(\pi^\pm \pi^0) <  0.475~\GEVcc$, which is below the $K^+$ mass considering the resolution of $0.003~\GEVcc$. 
 The latter process is suppressed by requiring the reconstructed 
$\mu^\pm \nu$ invariant mass to be  $m(\mu\nu) > 0.16~\GEVcc$, which is above the $\pi^+$  mass  considering the resolution of $0.004~\GEVcc$. 
Additionally, it is required that $m(\pi^\pm \pi^0) + p_{\pi^0,\perp}/c  < 0.6~\GEVcc$, 
where $p_{\pi^0,\perp}$ is the $\pi^0$ transverse momentum component with respect to the beam axis. The selection conditions,  illustrated in Fig. \ref{fig:km3cuts},  lead to 17\% signal loss  and reject 99.5\% of  the $K^\pm \to \pi^\pm \pi^0$ background.

Other background sources  considered are $K^\pm \to \pi^\pm \pi^0$ followed 
by $\pi^0  \to e^+ e^- \gamma$;  $K^\pm \to \pi^\pm \pi^0 \gamma$;  
 $K^\pm \to \pi^\pm \pi^0 \pi^0$ $(\pi^0 \to \gamma \gamma, ~\pi^0 \to e^+e^- \gamma)$;
$K^\pm \to  \pi^0 \pi^0 l^\pm   \nu$.
The $K^\pm _{\mu3}$ background to $K^\pm _{e3}$ decays arising from muon decay in flight is also 
considered. All these backgrounds  are   found to be negligible. 
The main background sources are summarized  in Table~\ref{tab:background}.  

\begin{table}[h]
  \renewcommand{\arraystretch}{1.1}
  \centering
  \begin{tabular}{| l | c | c |}
    \hline  
    Process  & $r_{e}$~[$10^{-3}$]    & $r_{\mu}$~[$10^{-3}$]                     \\ 
    \hline
    $K^\pm \to \pi^\pm \pi^0 \pi^0$ $(\pi^0 \to \gamma \gamma, \pi^0 \to \gamma \gamma)$   &             0.286(6)           &             2.192(32)          \\ 
    $K^\pm \to \pi^\pm \pi^0$ $(\pi^0 \to \gamma \gamma)$                                                 &            0.271(6)           &             0.392(10)          \\ 
    \hline
  \end{tabular}
  \caption{Background processes and 
   background to signal ratios  $r_{e}$ and $r_{\mu}$ in the selected $K^\pm _{e3}$ and $K^\pm _{\mu3}$ samples, estimated from MC simulations  described in  Section~\ref{MCsim}. The quoted errors include contributions from the external branching ratios and simulated statistics.
   } 
  \label{tab:background}
  \renewcommand{\arraystretch}{1.0}
\end{table}

\section{ Form factor measurement}
\label{sec:fit}
\begin{figure}[p]
\centering
\begin{minipage}{0.46\linewidth}
\includegraphics[width=1.\linewidth]{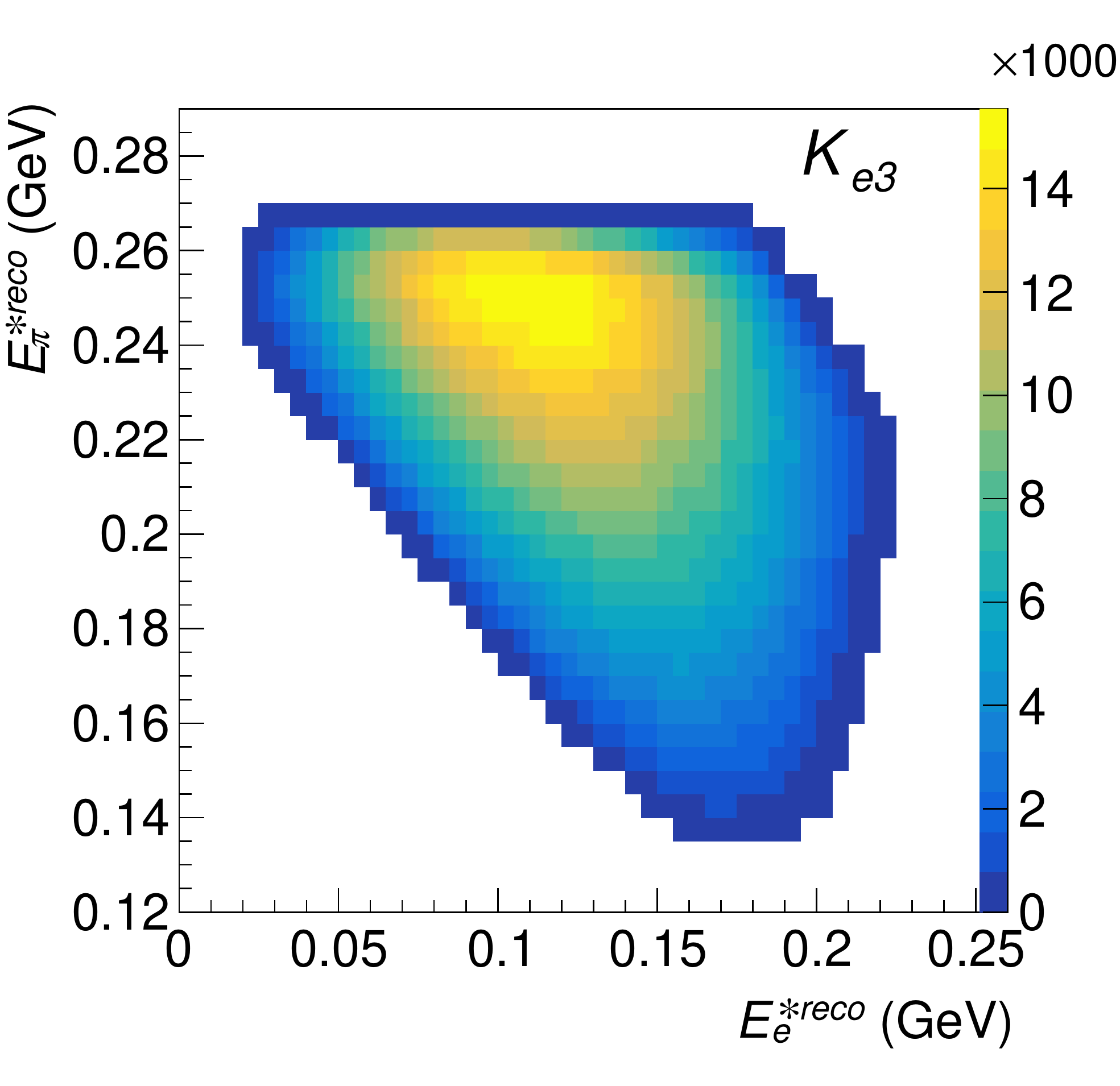}
\end{minipage} 
\begin{minipage}{0.46\linewidth}
\includegraphics[width=1.\linewidth]{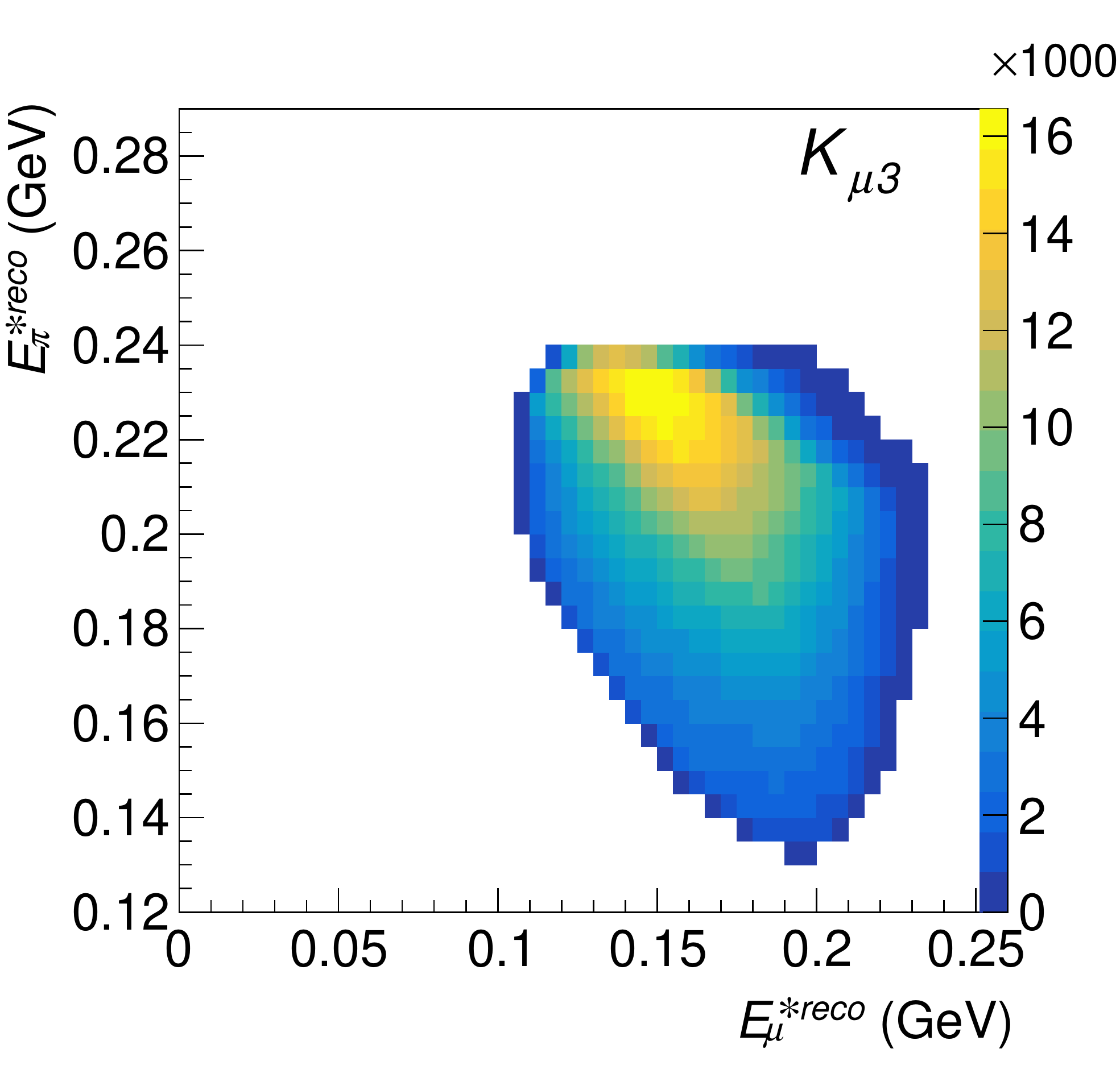}
\end{minipage}  
\\
\begin{minipage}{0.46\linewidth}
\includegraphics[width=1.\linewidth]{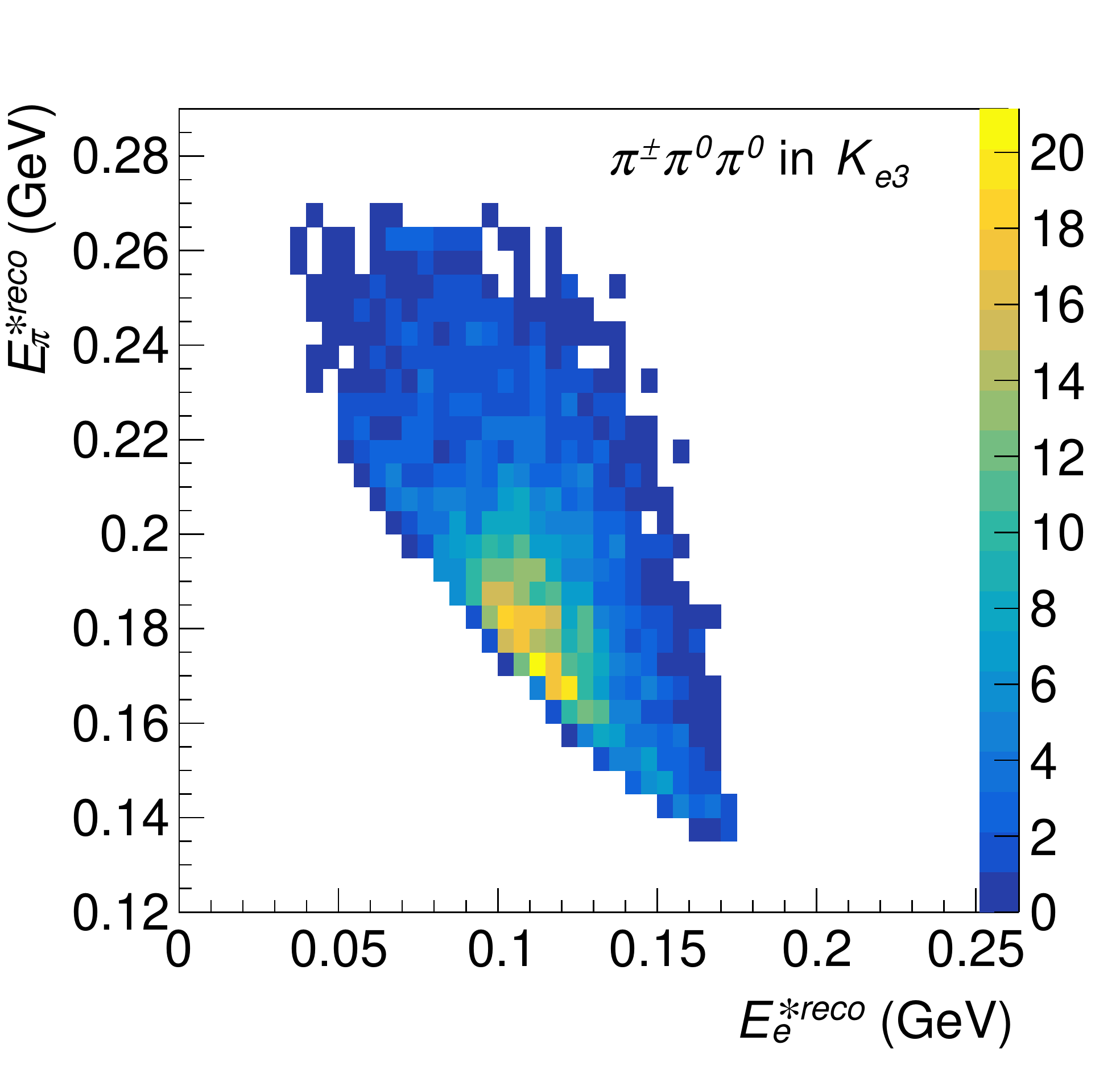}
\end{minipage} 
\begin{minipage}{0.46\linewidth}
\includegraphics[width=1.\linewidth]{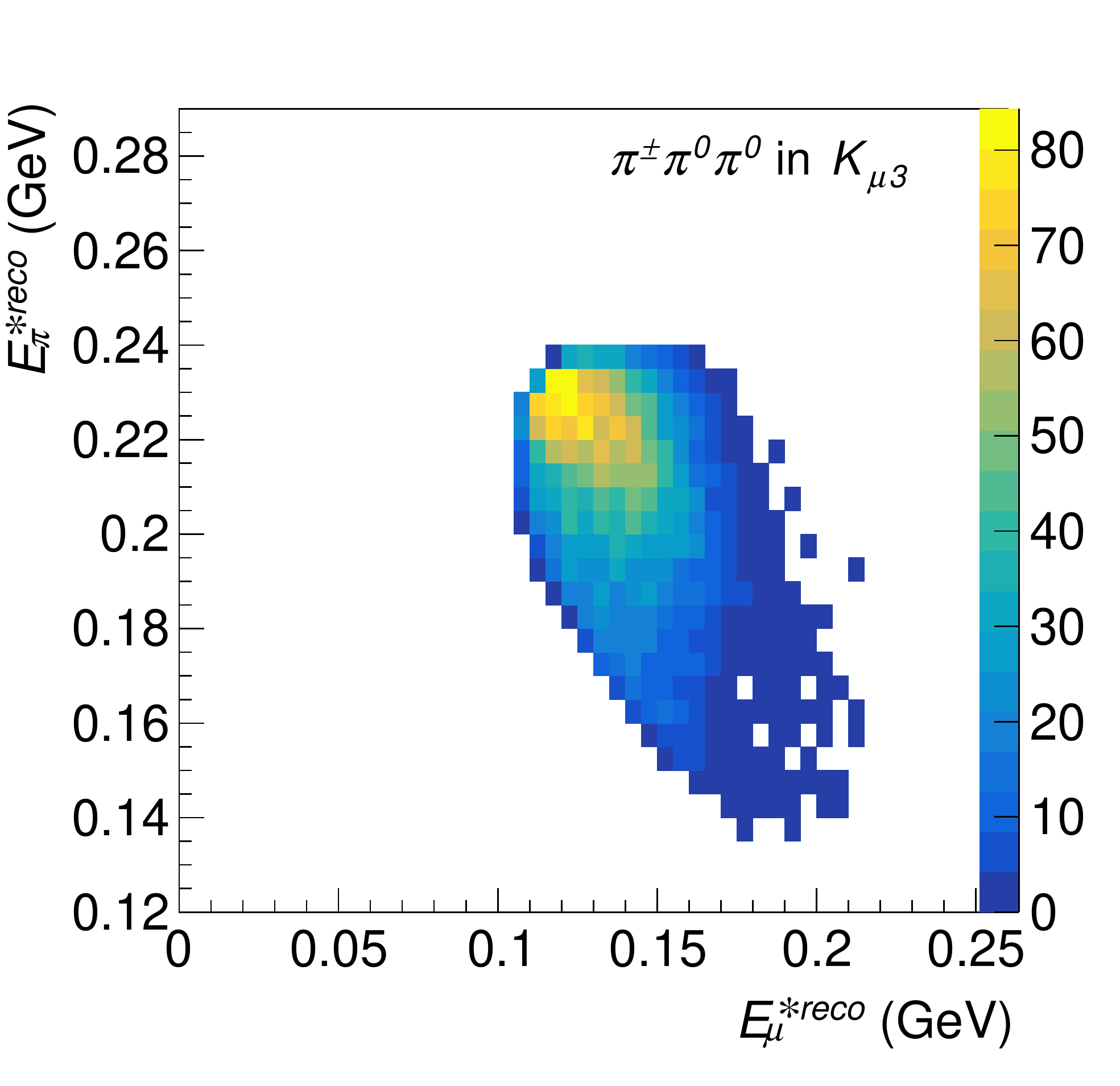}
\end{minipage}   
\\
\begin{minipage}{0.46\linewidth}
\includegraphics[width=1.\linewidth]{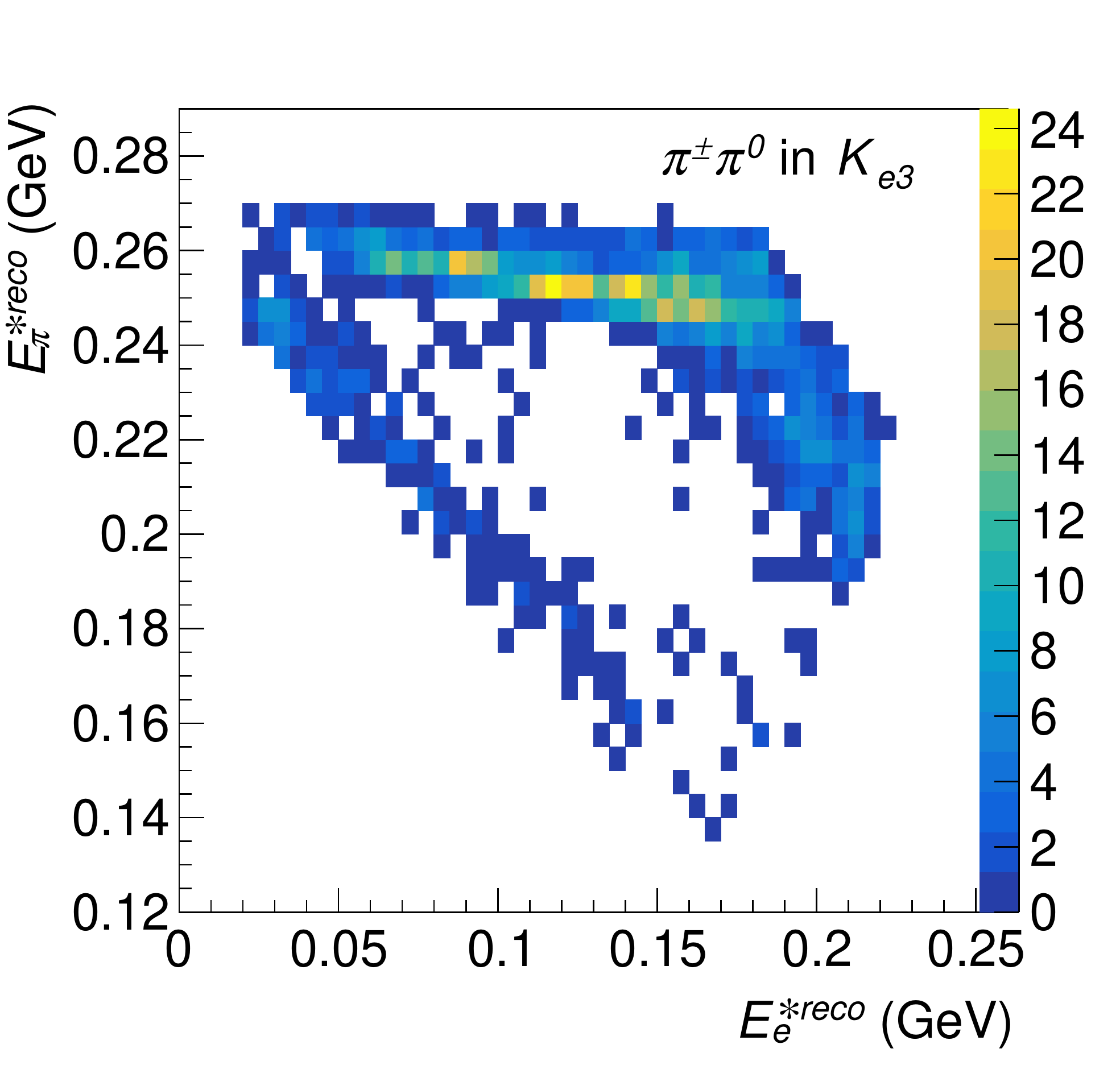}
\end{minipage} 
\begin{minipage}{0.46\linewidth}
\includegraphics[width=1.\linewidth]{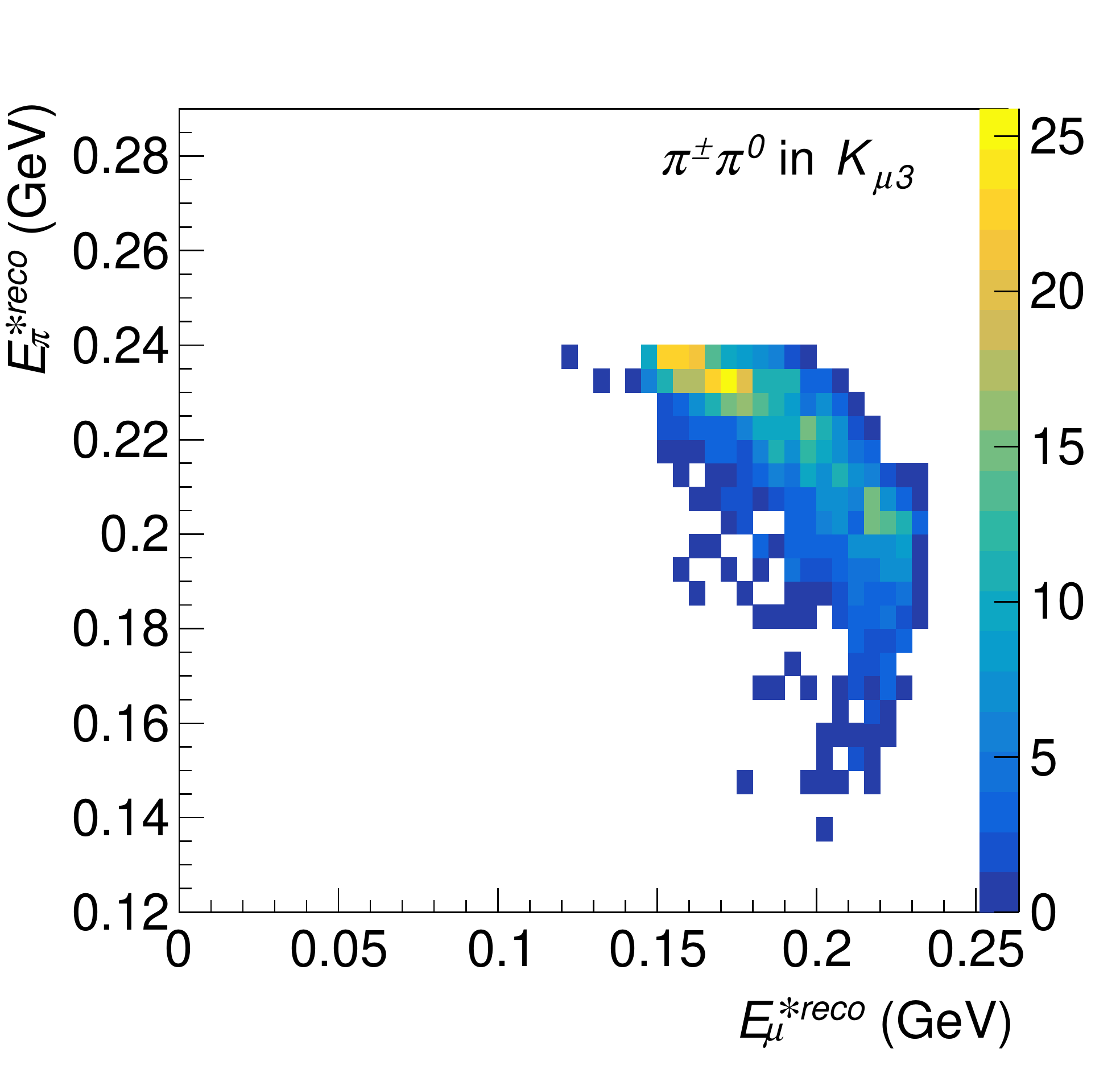}
\end{minipage} 
  \caption{Dalitz plot distributions  after the full selection of reconstructed $K^\pm _{l3}$ data events  (top row),
  simulated $K^\pm\to\pi^\pm\pi^0\pi^0$ (middle row) and  $K^\pm\to\pi^\pm\pi^0$  (bottom row) background events.
Left panels correspond to the $K^\pm _{e3}$  selection and  right panels to the $K^\pm _{\mu3}$ selection.
The simulated backgrounds are normalized to the total kaon flux in the data.  The cell size is  5 $\times$ 5~MeV$^2$. }
  \label{fig:edalitz}
\end{figure}

In total, $4.4  ~(2.3) \times 10^6$   reconstructed $K^\pm _{e3}$ ($K^\pm _{\mu3}$) candidates are selected from the data sample.
The Dalitz plot distributions, as defined in Eq.~(\ref{dalitz0}) and based on reconstructed energies,  are shown in Fig.~\ref{fig:edalitz} for the data and the main simulated backgrounds.

The FF parameters are measured independently for each of the two $K^\pm _{l3}$  decay modes.  A joint analysis is also performed by fitting simultaneously the two Dalitz plots  with a common set of FF parameters. 
 A set of FF parameters $\vec{\lambda}$  in each parameterization is measured by minimizing an estimator
  \begin{equation}
    \chi^2  (\vec{\lambda}, N) = \sum_{i} \frac{\,\left(\omega^\text{data}_i  - \omega^\text{bkg}_i (\vec{\lambda})-  N \cdot \omega^\text{sig}_i (\vec{\lambda}) \right)^2}{\sigma^2_{\omega^\text{data}_i}+ 
    \sigma^2_{\omega^\text{bkg}_i} (\vec{\lambda}) + N^2 \cdot \sigma^2_{\omega^\text{sig}_i} (\vec{\lambda})}, 
  \end{equation}
 where   the sum runs over all 5 $\times$ 5~MeV$^2$ Dalitz plot cells 
which have their  centres inside the kinematically allowed region of
non-radiative $K^\pm _{l3}$ events and contain at least 20 reconstructed data events.
Here $\omega^\text{data}_i$ is the population in cell $i$ of the reconstructed data Dalitz plot; 
$\omega^\text{sig}_i (\vec{\lambda})$  and $\omega^\text{bkg}_i (\vec{\lambda})$ are the  expected 
signal and background populations estimated from simulations;
$\sigma_{\omega_i ^\text{data}}$,   $\sigma_{\omega_i ^\text{sig}}$ and $\sigma_{\omega_i ^\text{bkg}}$ are the corresponding statistical errors;  
 $N$ is a normalization factor  that guarantees that the simulated sample is normalized to the data sample.

The quantities $\omega_i^\text{sig}(\vec{\lambda})$ are obtained at each iteration by applying a weight to each  simulated signal event, equal to the ratio of  the Dalitz plot density corresponding to  the parameter set $\vec{\lambda}$
 and the generated Dalitz plot density.  This approach accounts for the universal radiative corrections described in Section~\ref{MCsim}. The $\vec{\lambda}$-dependence of the background contribution
arises from  the dependence of the signal acceptances on the FFs.

\section{Systematic uncertainties}
\label{labelsystem}
The following sources of systematic uncertainties are considered. The resulting error estimates are assumed to be uncorrelated. 
\subsection{Experimental systematic uncertainties}

\paragraph{Beam  modelling} 
The diverging beam component which is not simulated (Section~\ref{sec:decayvertex})  gives rise to one of the largest systematic effects.  This effect is  evaluated by adding  specific samples of events, generated according to 
 the measured transverse beam profile,  to the simulated signal samples,  improving the Data/MC agreement of the $B$  spectra.
 The imperfect simulation of the kaon beam  spectrum leads to  variations  of the Data/MC ratio of reconstructed momentum spectra as a function of momentum within a few percent.
The corresponding systematic effect on the FF measurement is evaluated by assigning momentum-dependent  weights to the simulated events.
To evaluate the sensitivity of the results to the  beam average momentum value $p_B$  used in the selection (Section \ref{kaneumom}), which is reproduced by the MC simulation to a  precision of 0.03~$\GEVc$,
the analysis is repeated with the $p_B$  value shifted conservatively by $0.1~\GEVc$. 

\paragraph{LKr energy scale and non-linearity}
The $\pi^0$ reconstruction is sensitive to the LKr energy scale and non-linearities. 
The systematic uncertainty on the energy scale is 0.1\% (correlated between  data and simulated samples) while the energy scale difference between data and simulation is  known to 0.03\% precision. The systematic uncertainties on the FF measurement are estimated by varying the energy scale corrections 
within their uncertainties.
Cluster energies below 10 GeV are  affected by non-linearities in the energy scale.  
This  is  corrected  for,  and the residual systematic effects are estimated   by variation of the correction  method  as  detailed in \cite{Batley:2000zz}.

\paragraph{Residual background}
Systematic uncertainties on the background estimates 
are evaluated by studying the level of Data/MC agreement in background-enhanced control regions defined as $0.7<E/p<0.9$ for the $K^\pm _{e3}$  selection, and $B>15$
(corresponding to off-axis decay vertices, see Section~\ref{sec:decayvertex}) for the $K^\pm _{\mu 3}$ selection.
The  uncertainties assigned to background contributions  are $\delta r_e / r_e = 30\%$ and   $\delta r_{\mu} / r_{\mu} = 10\%$. They are propagated to the results, together with those listed in Table~\ref{tab:background}. 

\paragraph{Particle identification}
Electron identification efficiency  is determined by the lower $E/p$ condition. Using an almost background-free 
$K^\pm _{e3}$ data sample selected kinematically, 
 the efficiency  has been measured as a function of momentum to increase from 98\% at 5~$\GEVc$ to 99.6\%  above $10~\GEVc$.
 Efficiency measurements for data and simulated samples  agree  to better than 0.2\%.
Systematic uncertainties due to electron identification are evaluated by
weighting MC events to correct for  the residual  Data/MC  disagreement.
Muon identification inefficiency for $K^\pm _{\mu 3}$ decays is reduced  to the 0.1\% level, without dependence on the kinematic variables,
by the minimum muon momentum and  MUV geometrical acceptance requirements.  The corresponding systematic effect on the FF measurement is  negligible.

\paragraph{Event pileup} 
Pileup of  signal events with  independent kaon decays is not described by the simulation.  Effects of pileup  
are estimated by doubling the size of the maximum allowed time difference between 
the accepted photon candidates, and between  the accepted lepton and $\pi^0$ candidates. The shifts in 
the results are considered as systematic  uncertainties. 

\paragraph{Acceptance}
The Data/MC ratios of the decay vertex $z$ position  distributions (Fig.~\ref{fig:acceptance}) 
 reflect the quality of the acceptance simulation.  To account for the residual variation of these ratios, 
 the transverse cuts in  DCH, LKr and MUV detector planes  are widened by a factor of 1.002 in the  selection for the simulated  samples. The resulting variations of the FF parameters are considered as systematic  uncertainties. 
 
\paragraph{Neutrino momentum resolution} The cut on the squared longitudinal neutrino
momentum $\plnu^2$ is applied in the core region of the distribution  (Fig.~\ref{fig:pnu2neum}). 
A mismatch in $\plnu^2$ resolution between data and simulation  can therefore bias the results. 
Introducing an additional smearing for the simulated events, that is increasing the deviation of the reconstructed $\plnu^2$ from its true value by 1.5\%, leads to an improvement of the Data/MC agreement 
near the peak of the distribution.  The resulting variations are taken as corresponding systematic uncertainties

\paragraph{Trigger efficiency}
The trigger is based on uncorrelated HOD and LKr information  (Section~\ref{sec:beamdet}). 
Within the $K^\pm _{l3}$ selection, the  HOD trigger efficiency   is measured to be 0.9973(2) using a  control sample triggered 
by the NHOD, while the LKr trigger efficiency is measured to be 0.9987(1) using a 
control sample triggered by the HOD.
The total trigger efficiency  is obtained as the product of these two components. 
No statistically significant variations of the trigger efficiencies with the Dalitz plot variables  are observed.
Each efficiency component is measured as a function of  $\Ep$ and $\El$ variables and parameterized   with second order polynomial functions.   The statistical uncertainties on the parameters of these functions are propagated to the FF measurements, and the resulting variations considered as systematic uncertainties.

\paragraph{Dalitz plot binning and resolution}
The fit  has been repeated with  a 
Dalitz plot cell size reduced from 5 $\times$ 5~$\MEV^2$ to  2.5 $\times$ 2.5~$\MEV^2$. 
The resulting FF parameter variations  stay within the statistical errors.  However they are considered as systematic  uncertainties to account for
a possible imperfect description of the Dalitz plot density by the parameterizations.
To address the resolution effects, the FF measurement has been repeated using a different method, performing  a fit of the acceptance-corrected Dalitz plot by the density  function (\ref{dalitz0}).   Unlike the primary fit method, this procedure introduces a bias to the results due to  Dalitz plot resolution effects.
This bias is estimated by performing the same fit procedure  for  simulated signal samples with known input FF parameters replacing the data.
The  differences of the fit results between the two methods, corrected for the bias, are considered as  systematic uncertainties.

\subsection{External sources of systematics effects}

\paragraph{Radiative corrections}

The  FF parameters measured using the universal radiative corrections \cite{Cirigliano:2001mk} are not affected by theoretical uncertainties by construction. Nevertheless, for comparison with other
measurements and calculations,   the FF fits  have also been performed using radiative corrections  computed within the ChPT $e^2p^2$ approximation  \cite{Cirigliano:2001mk}.  The differences between the two sets of results are quoted  as external uncertainties.

\paragraph{External inputs} 

The  uncertainties on the  numerical inputs to the dispersive parameterization 
 (\ref{eq:gh})   are propagated to the FF fit results under the assumption that they are not correlated.

\section{Results}
\label{results}
\begin{figure}[ht]
  \begin{center}
    \includegraphics[width=\linewidth]{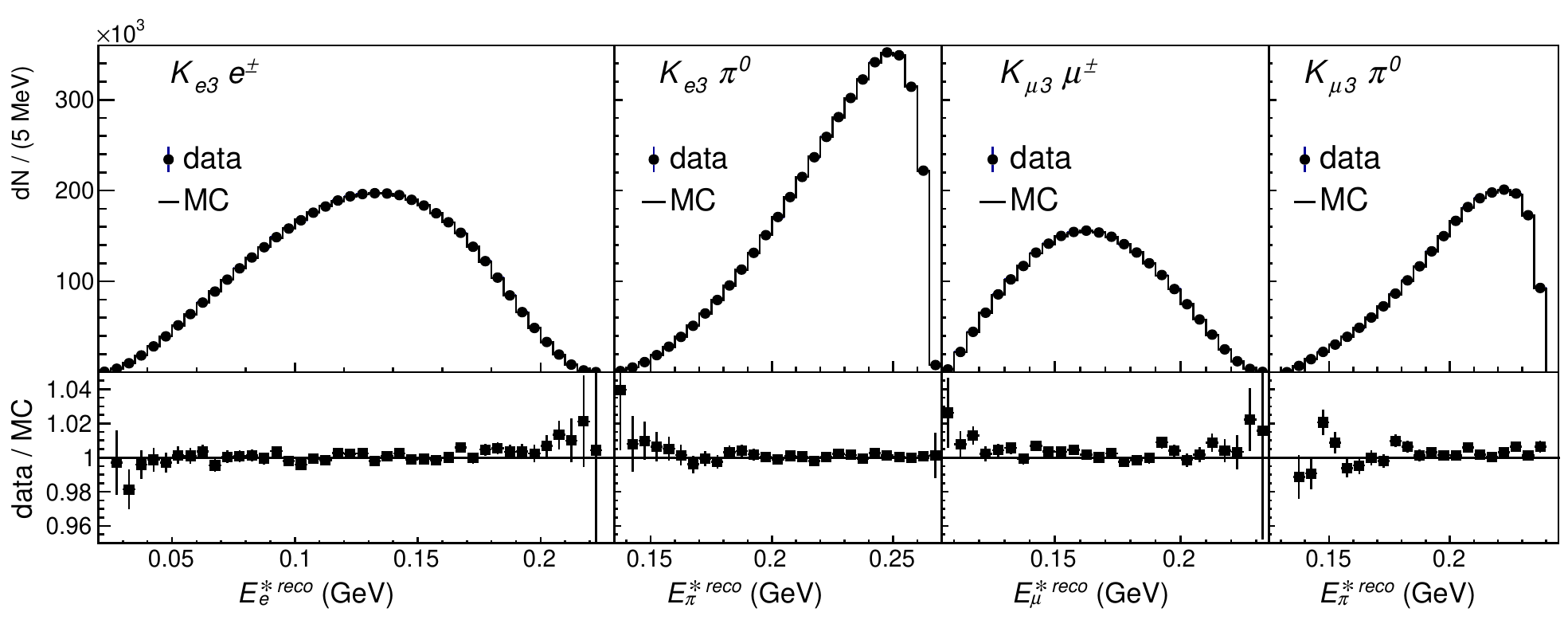}
  \end{center}
  \caption{Reconstructed lepton energy $\Elreco$  and pion energy $\Epreco$  distributions for 
  $K^\pm _{e3}$  and  $K^\pm _{\mu3}$   data  (after background subtraction)   and  simulated samples 
according to the fit results using the Taylor expansion model,  
  and corresponding  Data/MC ratios.   Simulated distributions according to fit results 
using other parameterizations cannot be distinguished within the resolution of the plots.}
  \label{fig:kl3fitpro}
\end{figure}

Lepton and pion energy 
projections of the reconstructed Dalitz plots for the data and the simulated samples corresponding to the  fit  results, along with  their ratios Data/MC, are shown in Fig.~\ref{fig:kl3fitpro}.
 The fit results are listed in Tables~\ref{tab:ke3fit}, \ref{tab:kmu3fit} and \ref{tab:jointfit} for
$K^\pm _{e3}$, $K^\pm _{\mu3}$ and the joint analysis, respectively.  
The  fit quality is satisfactory in all cases, as quantified by the $\chi^2$ values.
The quoted correlation coefficients are derived from sums of the covariance matrices 
of the statistical and the systematic uncertainties.
Form factor  measurements  from $K^\pm _{e3}$ and $K^\pm _{\mu3}$ decays are in agreement.   

\begin{figure}[ht]
 \begin{minipage}{0.32\linewidth}
  \includegraphics[width=1.\linewidth]{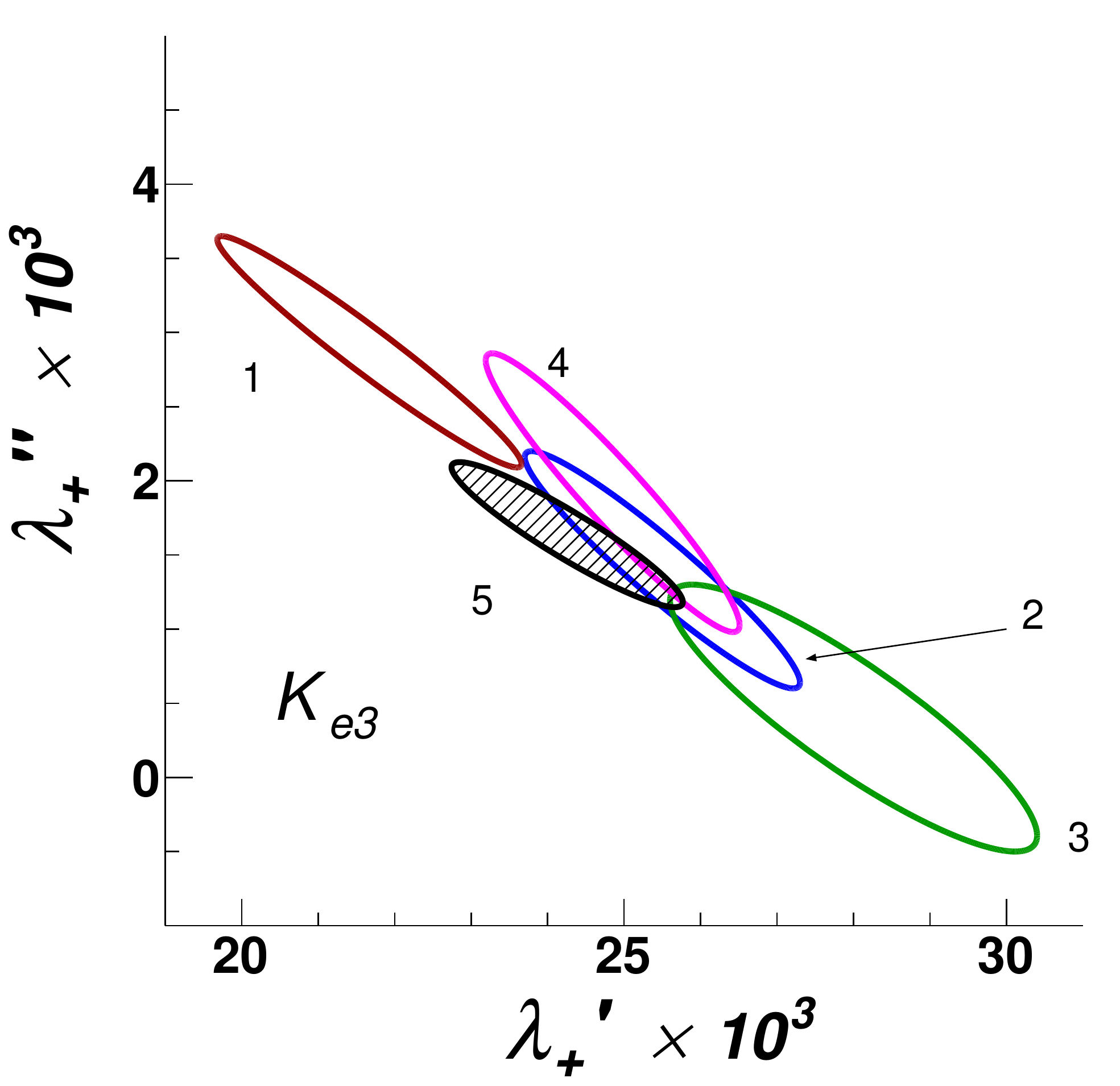}
\end{minipage} 
\begin{minipage}{0.32\linewidth}
\includegraphics[width=1.\linewidth]{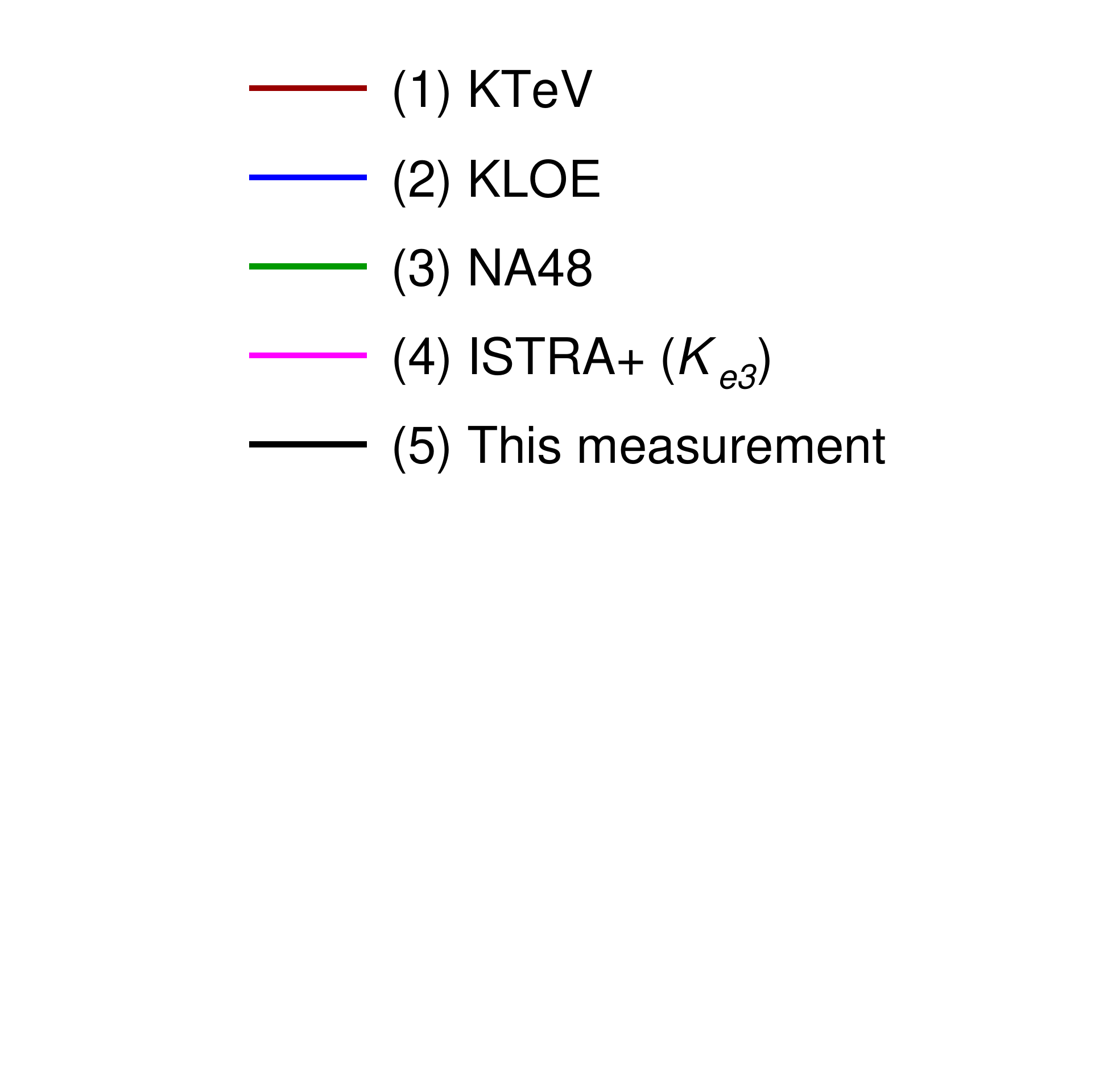} 
\end{minipage}   
\hspace{0.32\linewidth}\\
 \begin{minipage}{0.32\linewidth}
\includegraphics[width=1.\linewidth]{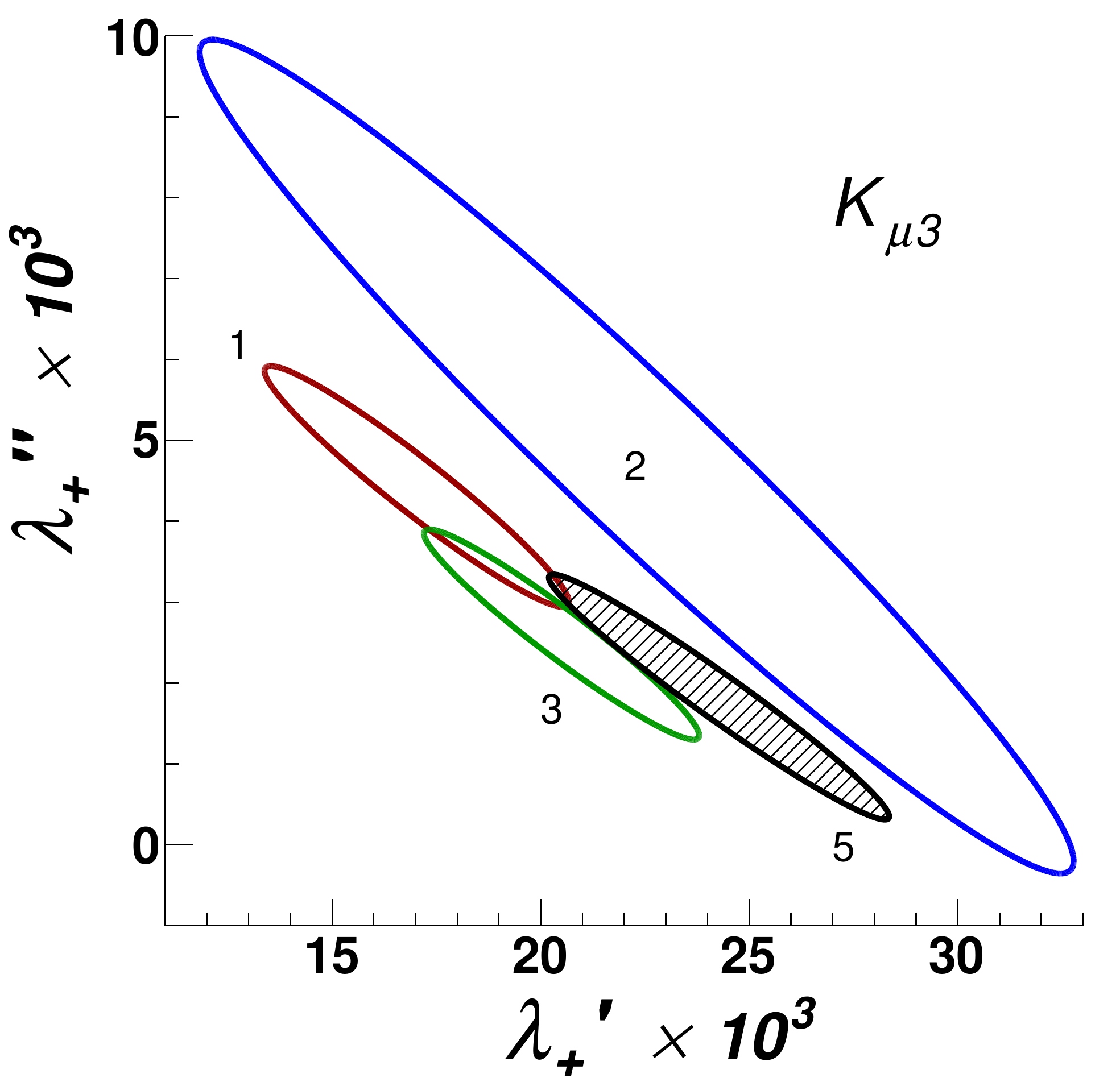}
\end{minipage} 
\begin{minipage}{0.32\linewidth}
\includegraphics[width=1.\linewidth]{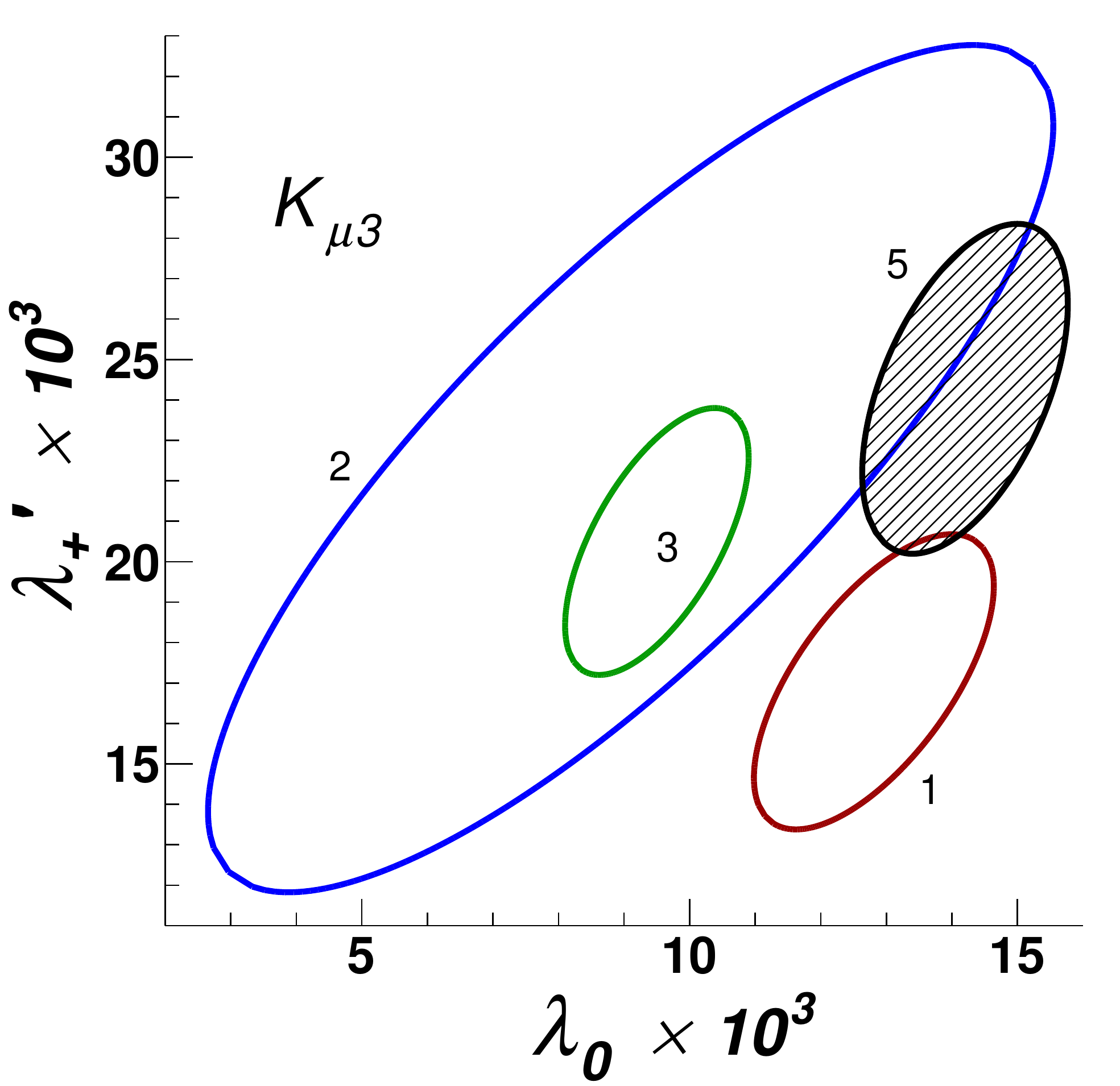} 
\end{minipage} 
\begin{minipage}{0.32\linewidth}
\includegraphics[width=1.\linewidth]{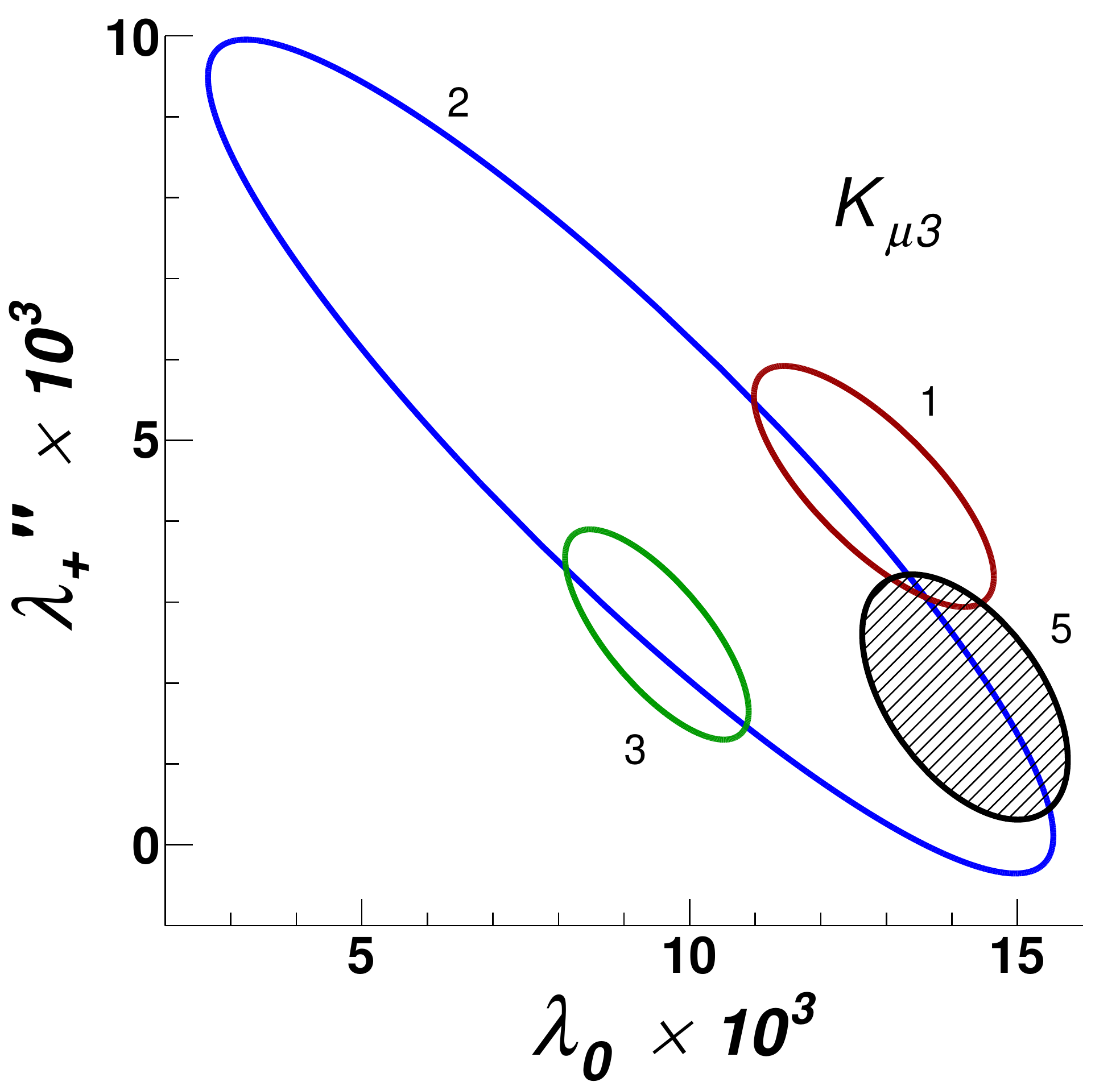} 
\end{minipage} 
  \caption{One sigma (39.4\% CL) contours for the obtained parameters of the Taylor expansion of the 
  $K_{e3}$ and  $K_{\mu3}$  FFs together with 
    measurements (obtained from  $K^0 _L$  or $K^-$ decays) by the KTeV~\cite{Alexopoulos:2004sy},
    KLOE~\cite{Ambrosino:2006gn,Ambrosino:2007ad},
    NA48~\cite{Lai:2004kb,Lai:2007dx}, and
    ISTRA+~\cite{Yushchenko:2003xz,Yushchenko:2004zs} Collaborations.  The  $K_{e3}$ results from NA48 and ISTRA+ have been modified by \cite{Antonelli:2010yf} to comply with the considered parameterization.  The  $K_{\mu3}$ results from ISTRA+ do not provide enough information to be displayed on the same panels as the other experimental results.} 
  \label{fig:km3ell3}
\end{figure}

\begin{figure}[ht]
 \begin{minipage}{0.32\linewidth}
\includegraphics[width=1.\linewidth]{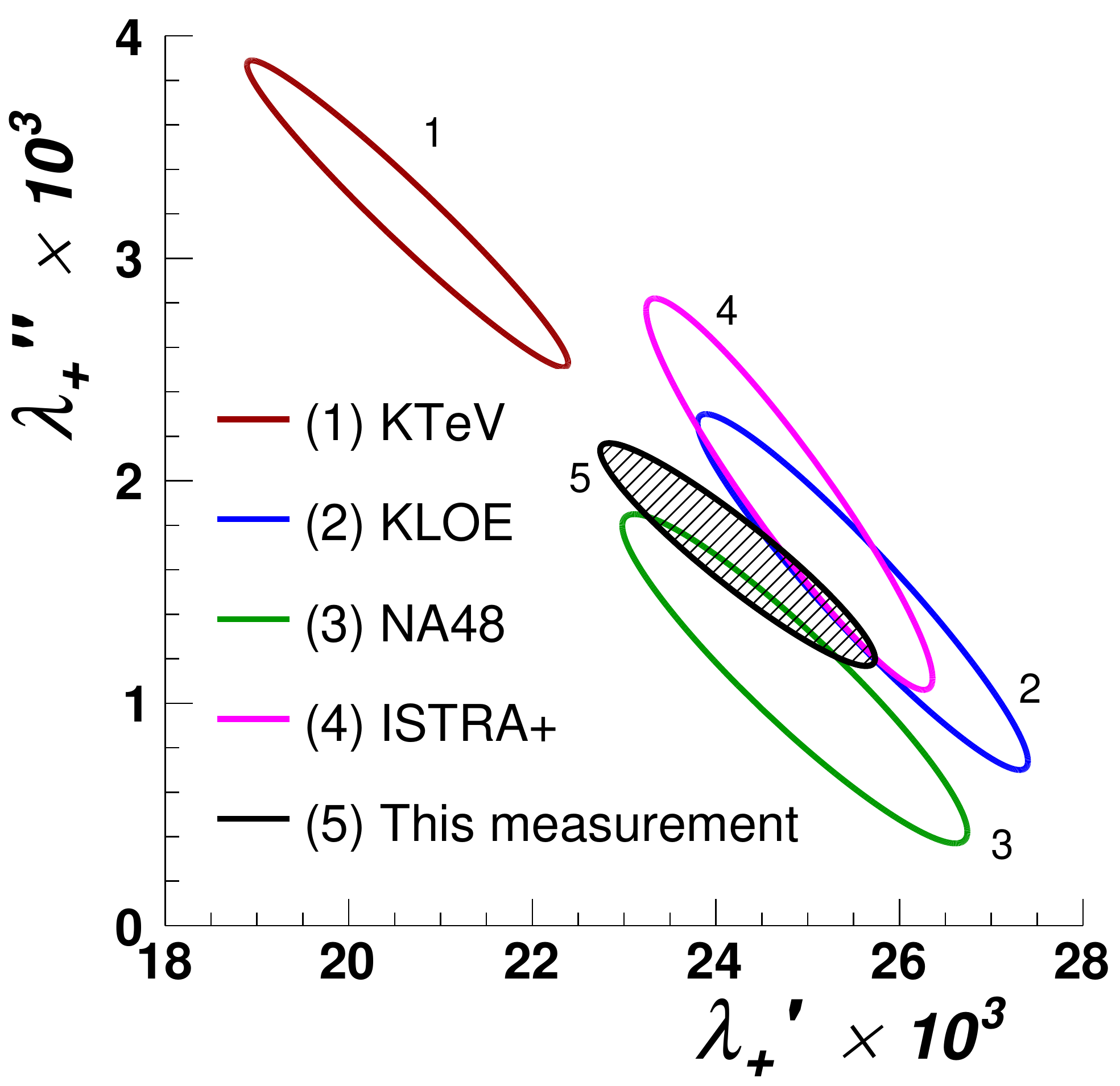}
\end{minipage} 
\begin{minipage}{0.32\linewidth}
\includegraphics[width=1.\linewidth]{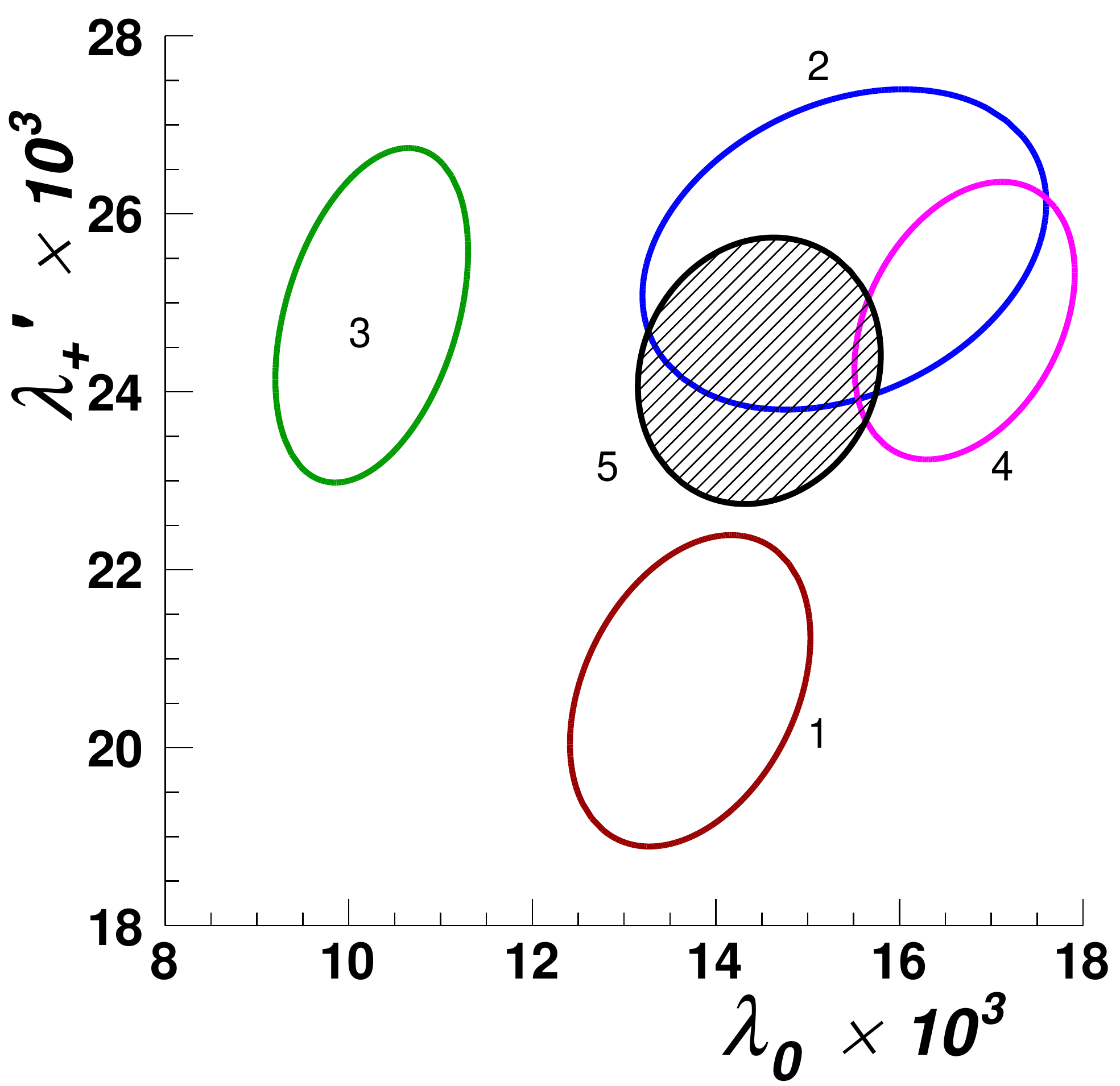} 
\end{minipage} 
\begin{minipage}{0.32\linewidth}
\includegraphics[width=1.\linewidth]{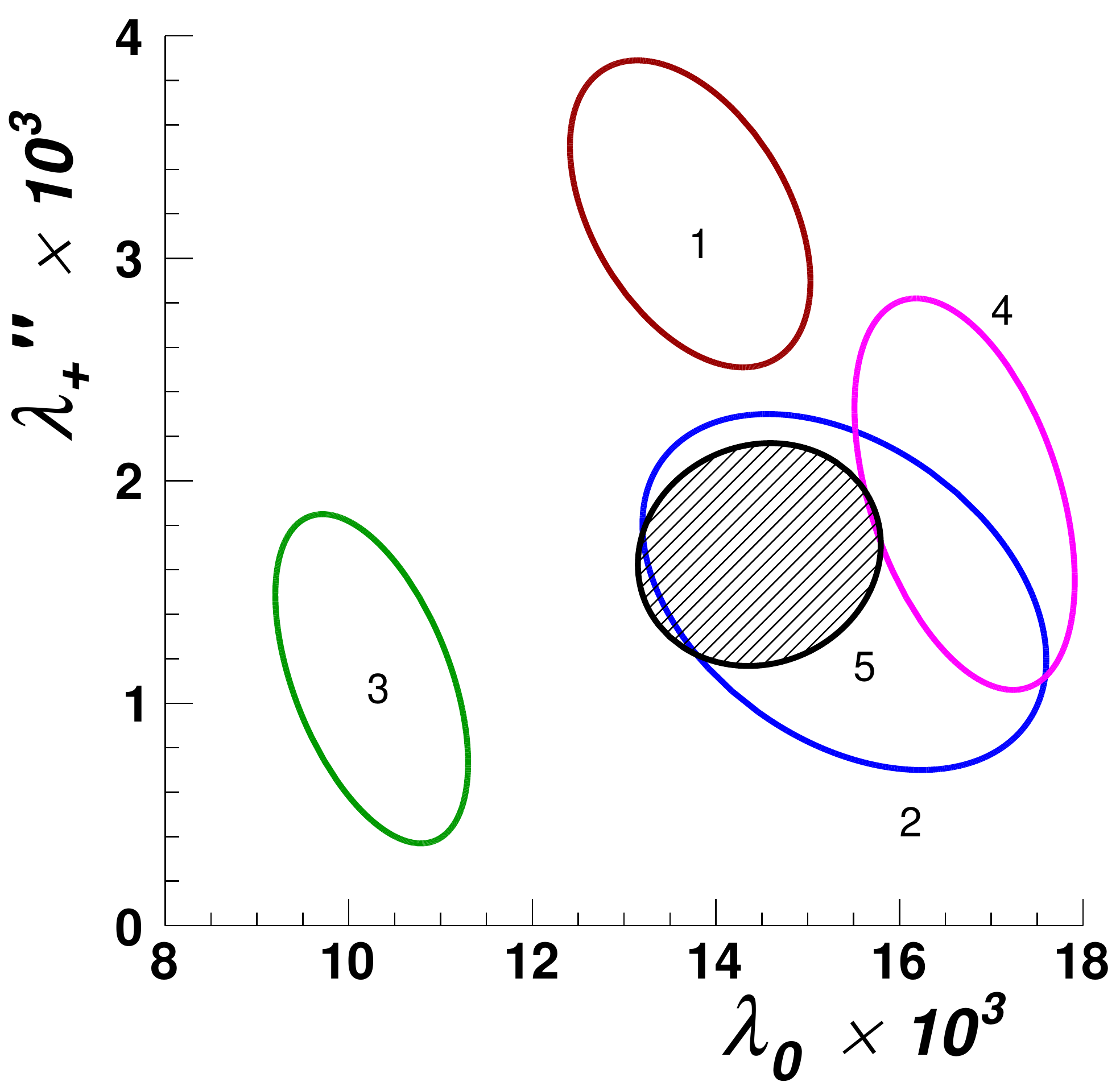} 
\end{minipage} 
 \caption{One sigma (39.4\% CL) contours for the  parameters of the Taylor 
expansion obtained from  the joint analysis  together with 
    the combinations of $K_{e3}$  and $K_{\mu 3}$ measurements  by the KTeV~\cite{Alexopoulos:2004sy},
    KLOE~\cite{Ambrosino:2006gn,Ambrosino:2007ad},
    NA48~\cite{Lai:2004kb,Lai:2007dx}, and
    ISTRA+~\cite{Yushchenko:2003xz,Yushchenko:2004zs} Collaborations provided by~\cite{Antonelli:2010yf}.}
  \label{fig:kl3ell3}
\end{figure}

The results of the present analysis for the Taylor expansion parameterization,  together 
with the earlier  results from
KTeV~\cite{Alexopoulos:2004sy},
KLOE~\cite{Ambrosino:2006gn,Ambrosino:2007ad},
NA48~\cite{Lai:2004kb,Lai:2007dx}, and
ISTRA+~\cite{Yushchenko:2003xz,Yushchenko:2004zs} experiments, as reviewed in
\cite{Antonelli:2010yf}, are shown in Fig.~\ref{fig:km3ell3}, \ref{fig:kl3ell3}. The present results are in agreement with the previous measurements and have similar or better precision. 
\begin{table}[!ht]
  \centering
  \begin{tabular}{|l|rr|c|c|} \hline 
    & {$\lambda'_+$} & {$\lambda''_+$} & {$m_V$} & {$\Lambda_+$} \\ \hline
 \textbf{Central values} & 24.26 & 1.64   & 885.2   & 24.94 \\ \hline
 \textbf{Statistical error } & 0.78 & 0.30   & 3.3   & 0.21 \\ \hline
 Diverging beam component & 0.89 & 0.31  & 1.4  & 0.10 \\
 Kaon momentum spectrum  & 0.00 & 0.00  & 0.1  & 0.01 \\
 Kaon mean momentum  & 0.03 & 0.01  & 0.1  & 0.01 \\
 LKr~energy scale  & 0.69 & 0.14  & 5.0  & 0.33 \\
LKr~non-linearity & 0.28 & 0.01  & 3.4  & 0.22 \\
 Residual background & 0.08 & 0.04  & 0.4  & 0.02 \\
 Electron identification & 0.02 & 0.01  & 0.2  & 0.01 \\
 Event pileup & 0.24 & 0.08  & 0.5  & 0.03 \\
 Acceptance & 0.29 & 0.08  & 1.2  & 0.08 \\
  Neutrino momentum resolution & 0.18 & 0.04  & 1.1  & 0.07 \\
 Trigger efficiency & 0.33 & 0.13  & 1.0  & 0.07 \\
  Dalitz plot binning & 0.07 & 0.01  & 0.7  & 0.05 \\
 Dalitz plot resolution & 0.06 & 0.04  & 0.4  & 0.02 \\
 \hline
Radiative corrections  & 0.20 & 0.01  & 2.9  & 0.19 \\
 External inputs  & & & & 0.44 \\
     \hline
 \textbf{Systematic error} & 1.30 & 0.39   & 7.2   & 0.64 \\ \hline
 \textbf{Total error} & 1.51 & 0.49   & 7.9   & 0.67 \\ \hline
    Correlation coefficient      & \multicolumn{2}{c|}{$-\,0.929$} & {---} & {---} \\ \hline
    $\chi^2$/NDF & \multicolumn{2}{c|}{569.1/687} & {568.9/688} & {569.0/688} \\ \hline
  \end{tabular}
\caption{Form factor results of the $K^\pm _{e3}$ analysis. The correlation includes both statistical and systematic uncertainties.
 The units of $\lambda'_+$, $\lambda''_+$ and $\Lambda_+$ values and errors are $10^{-3}$. The units of $m_V$  value and error are $\MEVcc$.
}
\label{tab:ke3fit} 
\end{table}

\begin{table}[!ht]
  \centering
  \begin{tabular}{|l|rrr|
                     rr|
                     rr|} \hline
    & {$\lambda'_+$} & 
    $\lambda''_+$ & {$\lambda_0$}
         & {$m_V$} & {$m_S$} & {$\Lambda_+$} & {$\ln C$} \\ \hline
 \textbf{Central values} & 24.27 & 1.83 & 14.20   & 878.4 & 1214.8   & 25.36 & 182.17 \\ \hline
 \textbf{Statistical error } & 2.88 & 1.05 & 1.14   & 8.8 & 23.5   & 0.58 & 6.31 \\ \hline
 Diverging beam component & 2.03 & 0.78 & 0.13    & 0.9 & 30.9    & 0.04 & 8.98 \\  
 Kaon momentum spectrum  & 0.08 & 0.02 & 0.00    & 0.1 & 0.9    & 0.01 & 0.24 \\  
 Kaon mean momentum & 0.06 & 0.00 & 0.06    & 0.8 & 2.4    & 0.06 & 0.63 \\  
 LKr~energy scale  & 0.31 & 0.01 & 0.53  & 4.5 & 19.4  & 0.30 & 5.55 \\
 LKr~non-linearity & 0.93 & 0.38 & 0.25    & 1.3 & 21.7    & 0.08 & 6.26 \\  
 Residual background & 0.13 & 0.00 & 0.02    & 1.7 & 1.3    & 0.11 & 0.31 \\  
 Event pileup & 0.04 & 0.01 & 0.03    & 0.0 & 0.7    & 0.00 & 0.18 \\  
 Acceptance & 0.70 & 0.18 & 0.18    & 2.9 & 0.3    & 0.20 & 0.14 \\  
 Neutrino momentum resolution & 0.09 & 0.03 & 0.08    & 0.2 & 2.1    & 0.01 & 0.59 \\  
 Trigger efficiency & 0.60 & 0.08 & 0.23    & 5.1 & 5.7    & 0.35 & 1.72 \\  
 Dalitz plot binning & 1.50 & 0.63 & 0.63    & 2.8 & 3.6    & 0.18 & 0.85 \\  
 Dalitz plot resolution & 0.04 & 0.01 & 0.02    & 0.1 & 0.4    & 0.01 & 0.18 \\  
 \hline
 Radiative corrections  & 0.32 & 0.10 & 0.54  & 0.7 & 23.7  & 0.04 & 6.73 \\
 External inputs  & & & & & & 0.46 & 2.87 \\  
     \hline
 \textbf{Systematic error} & 2.89 & 1.09 & 1.07   & 8.3 & 49.2   & 0.72 & 14.45 \\ \hline
 \textbf{Total error} & 4.08 & 1.52 & 1.57   & 12.1 & 54.5   & 0.92 & 15.76 \\ \hline
 Correlation coefficients  & \multicolumn{3}{c|}{$-0.974$~($\lambda'_+/\lambda''_+$) } 
                            & \multicolumn{2}{c|}{$0.029$} & \multicolumn{2}{c|}{$0.104$} \\ 
                            & \multicolumn{3}{c|}{$\phantom{-}0.511$~($\lambda'_+/\lambda_0$)} & & &&\\
                            & \multicolumn{3}{c|}{$-0.513$~($\lambda''_+/\lambda_0$)} & & & & \\ \hline 
    $\chi^2$/NDF  & \multicolumn{3}{c|}{409.9/381} & \multicolumn{2}{c|}{409.9/382} & \multicolumn{2}{c|}{410.3/382}\\ \hline
  \end{tabular}
  \caption{Form factor results of the $K^\pm _{\mu3}$ analysis. The correlations include both statistical and systematic uncertainties. 
           The units of $\lambda'_+$, $\lambda''_+$, $\lambda_0$, $\Lambda_+$ and $\ln C$ values and errors are $10^{-3}$. The units of $m_V$ and  $m_S$ values and errors are $\MEVcc$.
 } \label{tab:kmu3fit}
\end{table}

\begin{table}[!ht]
  \centering
  \begin{tabular}{|l|rrr|
                     rr|
                     rr|} \hline
    & {$\lambda'_+$} & {
    $\lambda''_+$} & {$\lambda_0$}
         & {$m_V$} & {$m_S$} & {$\Lambda_+$} & {$\ln C$} \\ \hline
 \textbf{Central values} & 24.24 & 1.67 & 14.47   & 884.4 & 1208.3   & 24.99 & 183.65 \\ \hline
 \textbf{Statistical error} & 0.75 & 0.29 & 0.63   & 3.1 & 21.2   & 0.20 & 5.92 \\ \hline
  Diverging beam component & 0.97 & 0.35 & 0.55    & 1.1 & 32.2    & 0.08 & 9.43 \\  
 Kaon momentum spectrum  & 0.00 & 0.00 & 0.02    & 0.1 & 0.7    & 0.00 & 0.19 \\  
 Kaon mean momentum  & 0.04 & 0.01 & 0.04    & 0.2 & 1.7    & 0.01 & 0.47 \\  
 LKr~energy scale  & 0.66 & 0.12 & 0.61  & 4.9 & 17.4  & 0.32 & 5.16 \\
 LKr~non-linearity & 0.20 & 0.01 & 0.55    & 3.1 & 19.6    & 0.20 & 5.77 \\  
 Residual background & 0.08 & 0.03 & 0.04    & 0.1 & 0.7    & 0.01 & 0.16 \\  
 Electron identification  & 0.01 & 0.01 & 0.01    & 0.2 & 0.2    & 0.01 & 0.05 \\  
 Event pileup & 0.23 & 0.08 & 0.08    & 0.4 & 0.2    & 0.03 & 0.07 \\  
 Acceptance & 0.23 & 0.07 & 0.03    & 0.7 & 4.3    & 0.05 & 1.11 \\  
 Neutrino momentum resolution & 0.16 & 0.04 & 0.04    & 0.9 & 3.3    & 0.06 & 0.88 \\  
 Trigger efficiency & 0.29 & 0.13 & 0.20    & 1.1 & 9.9    & 0.07 & 2.82 \\  
 Dalitz plot binning & 0.05 & 0.04 & 0.06    & 0.9 & 1.1    & 0.06 & 0.29 \\  
 Dalitz plot resolution & 0.02 & 0.01 & 0.03    & 0.0 & 1.3    & 0.00 & 0.39 \\  
\hline
 Radiative corrections  & 0.17 & 0.01 & 0.57  & 2.5 & 20.1  & 0.16 & 5.92 \\
External inputs   & & & & & & 0.44 & 2.94 \\  
     \hline
 \textbf{Systematic error} & 1.30 & 0.41 & 1.17   & 6.7 & 47.5   & 0.62 & 14.25 \\ \hline
 \textbf{Total error} & 1.50 & 0.50 & 1.32   & 7.4 & 52.1   & 0.65 & 15.43 \\ \hline
    Correlation coefficient   & \multicolumn{3}{c|}{$-0.934$~($\lambda'_+/\lambda''_+$)} 
                           & \multicolumn{2}{c|}{$0.374$} & \multicolumn{2}{c|}{$0.354$} \\ 
                           & \multicolumn{3}{c|}{ \phantom{9}$0.118$~($\lambda'_+/\lambda_0$)} & & &&\\
                           & \multicolumn{3}{c|}{$\phantom{-}0.091$~($\lambda''_+/\lambda_0$)} & & & & \\ \hline 
    $\chi^2$/NDF & \multicolumn{3}{c|}{979.6/1070} & \multicolumn{2}{c|}{979.3/1071} & \multicolumn{2}{c|}{979.7/1071} \\ \hline
  \end{tabular}
  \caption{Form factor results of the joint $K^\pm _{l3}$ analysis. The correlations include both statistical and systematic uncertainties. 
  The units of $\lambda'_+$, $\lambda''_+$, $\lambda_0$, $\Lambda_+$ and $\ln C$ values and errors are $10^{-3}$. The units of $m_V$ and  $m_S$ values and errors are $\MEVcc$.
 } \label{tab:jointfit}
\end{table}


\clearpage
\section*{Acknowledgements}
We gratefully acknowledge the CERN SPS accelerator and beam line staff
for the excellent performance of the beam and the technical staff of
the participating institutes for their efforts in the maintenance and
operation of the detector, and data processing. We are grateful 
to Matthew Moulson for useful discussions.


\begin{thebibliography}{99}
 \bibitem{Batley:2007aa} 
 J.R.~Batley  {\em et al.}  (NA48/2 Collaboration),  Eur.\ Phys.\ J.\ {\textbf C}  {\boldmath $52$} (2007) 875.

\bibitem{Antonelli:2010yf}
 M.~Antonelli {\em et al.}  (FlaviaNet Working Group on Kaon Decays),
Eur.\ Phys.\ J.\ {\textbf C}  {\boldmath $69$} (2010) 399.

\bibitem{Chounet:1971yy}
L.M.~Chounet, J.-M.~Gaillard, M.-K.~Gaillard,
Phys.~Rept.  {\boldmath $4$} (1972) 199.

\bibitem{Olive:2016xmw}
M.~Tanabashi {\em et al.}  (Particle Data Group), Phys. Rev. {\textbf  D}  {\boldmath $98$} (2018) 030001.

\bibitem{Dennery:1963}
P.~Dennery,  H.~Primakoff,
 Phys.~Rev.  {\boldmath $131$} (1963) 1334.

\bibitem{Lichard:1997ya}
P.~Lichard, 
 Phys.~Rev.  {\textbf D} {\boldmath $55$} (1997) 5385.

\bibitem{Bernard:2009zm}
V.~Bernard, M.~Oertel, E.~Passemar, J.~Stern, 
 Phys.~Rev.  {\textbf D} {\boldmath $80$} (2009) 034034.

\bibitem{Fanti:2007vi}
V.~Fanti  {\em et al.}  (NA48 Collaboration),  Nucl. Instrum. Methods {\textbf A}  {\boldmath $574$} (2007) 443.

 \bibitem{Brun:1978fy}
 GEANT {\em detector description and simulation tool}, CERN program library long writeup {\textbf W{}\boldmath $5013$}, CERN, Geneva, Switzerland (1994).     

\bibitem{Gatti:2005kw}
C.~Gatti, Eur.\ Phys.\ J.\ {\textbf C}  {\boldmath $45$} (2006) 417.

\bibitem{Weinberg:1965nx}
S.~Weinberg, 
 Phys. Rev.  {\boldmath $140$} (1965) B516.

\bibitem{KTEV}
T. ~Alexopoulos {\em et al.} (KTEV Collaboration), Phys. Rev. {\textbf D} {\boldmath $71$}  (2005) 012001.

\bibitem{NA48}
A. ~Lai {\em et al.} (NA48 Collaboration), Phys. Lett.  {\textbf B} {\boldmath $605$} (2005) 247.

\bibitem{Cirigliano:2001mk}
V.~Cirigliano,  M.~Knecht,  H.~Neufeld, H.~Rupertsberger, P.~Talavera,
Eur.\ Phys.\ J.\ {\textbf C}  {\boldmath $23$} (2002) 121.

 \bibitem{Batley:2000zz} 
 J.R.~Batley  {\em et al.}  (NA48/2 Collaboration),  Eur.\ Phys.\ J.\ {\textbf C}  {\boldmath $64$} (2009) 589.

\bibitem{Alexopoulos:2004sy}
T.~Alexopoulos {\em et al.} (KTeV Collaboration), 
 Phys.~Rev. {\textbf D} {\boldmath $70$} (2004) 092007.

\bibitem{Ambrosino:2006gn}
F.~Ambrosino {\em et al.}  (KLOE Collaboration), Phys.~Lett. {\textbf B} {\boldmath $636$} (2006) 166.


\bibitem{Ambrosino:2007ad}
F.~Ambrosino  {\em et al.}  (KLOE Collaboration), JHEP {\boldmath $0712$} (2007) 105.

\bibitem{Lai:2004kb}
A.~Lai  {\em et al.} (NA48 Collaboration), Phys.~Lett. {\textbf B} {\boldmath $604$} (2004) 1.

\bibitem{Lai:2007dx}
A.~Lai  {\em et al.} (NA48 Collaboration), Phys.~Lett. {\textbf B} {\boldmath $647$} (2007) 341.


\bibitem{Yushchenko:2003xz}
O.~Yushchenko {\em et al.}, (ISTRA+ Collaboration), Phys.~Lett. {\textbf B} {\boldmath $581$} (2004) 31.
 
\bibitem{Yushchenko:2004zs}
O.~Yushchenko {\em et al.}, (ISTRA+ Collaboration), Phys.~Lett. {\textbf B} {\boldmath $589$} (2004) 111.


\end{thebibliography}
\end{document}